\documentclass[aps,prl,longbibliography,twocolumn,superscriptaddress,amsfont,graphicx,nofootinbib,preprintnumbers,10pt]{revtex4-1}%
\UseRawInputEncoding
\usepackage{ifpdf}
\usepackage{amsmath}
\usepackage{bm}
\usepackage[english]{babel}
\usepackage{amssymb}
\usepackage{braket}
\usepackage{setspace}
\usepackage{booktabs}
\usepackage{fontawesome}
\usepackage{afterpage}
\usepackage{booktabs} 
\allowdisplaybreaks[4]

\usepackage{hyperref}
\usepackage{enumerate}
\usepackage{lipsum}

\usepackage{slashed}

\usepackage{changes}

\begin{document}

\title{Search for High-Frequency Gravitational Waves via Geomagnetic Conversion with Radio Telescopes}

\author{Hongliang Tian}
\email{hongliangtian@njnu.edu.cn}
\affiliation{Department of Physics and Institute of Theoretical Physics, Nanjing Normal University, Nanjing, 210023, China}

\author{Lei Wu}
\email{leiwu@njnu.edu.cn}
\affiliation{Department of Physics and Institute of Theoretical Physics, Nanjing Normal University, Nanjing, 210023, China}
\affiliation{Nanjing Key Laboratory of Particle Physics and Astrophysics, Nanjing, 210023, China}

\author{Xiaolong Yang}
\email{yangxl@shao.ac.cn}
\affiliation{Shanghai Astronomical Observatory, Chinese Academy of Sciences, Shanghai 200030, China}
\affiliation{Shanghai Key Laboratory of Space Navigation and Positioning Techniques, Shanghai 200030, China}

\author{Qiang Yuan}
\email{yuanq@pmo.ac.cn}
\affiliation{Key Laboratory of Dark Matter and Space Astronomy, Purple Mountain Observatory, Chinese Academy of Sciences, Nanjing 210023, China}
\affiliation{School of Astronomy and Space Science, University of Science and Technology of China, Hefei 230026, China}

\author{Bin Zhu}
\email{zhubin@mail.nankai.edu.cn}
\affiliation{School of Physics, Yantai University, Yantai 264005, China}

\date{\today}

\begin{abstract}
The detection of high-frequency gravitational waves (HFGWs) above 10~kHz provides a crucial probe of exotic astrophysical phenomena and new physics. We report the first search for HFGWs via their conversion to electromagnetic radiation through the inverse Gertsenshtein effect in Earth's magnetic field, utilizing radio telescopes including the Very Large Array (VLA) and the Atacama Large Millimeter/submillimeter Array (ALMA). Since no statistically significant signal is observed, we obtain new upper limits on the characteristic strain across the 1~GHz -- 1~THz band, with the most stringent constraint reaching $h_c \lesssim 10^{-18}$, improving upon existing bounds by up to three orders of magnitude. These results significantly advance the exploration of uncharted parameter space for exotic gravitational-wave sources, paving the way for future discoveries with next-generation facilities such as the Square Kilometre Array (SKA).
\end{abstract}

\maketitle

{\it Introduction.}~~ Gravitational waves, a cornerstone prediction of general relativity, were first detected by LIGO in 2015, inaugurating a new observational window onto the cosmos \cite{LIGOScientific:2016aoc}. To date, detections have been confined to frequencies of hundreds of hertz, primarily from merging stellar-mass black holes and neutron stars \cite{LIGOScientific:2016aoc,LIGOScientific:2016sjg,LIGOScientific:2017vwq,LIGOScientific:2017zic,LIGOScientific:2018mvr,LIGOScientific:2020ibl,KAGRA:2021vkt}. Gravitational waves at other frequencies remain elusive, despite sustained efforts spanning from nanohertz frequencies with pulsar timing arrays \cite{Hobbs:2009yy,Manchester:2012za,McLaughlin:2013ira,Janssen:2014dka,EPTA:2015qep,Smith:2006nka,Pagano:2015hma,Clarke:2020bil,NANOGrav:2023gor,Xu:2023wog,Reardon:2023gzh,EPTA:2023fyk,Miles:2024seg} to high-frequency regimes with laboratory experiments \cite{Domcke:2022rgu,Ejlli:2019bqj,Holometer:2016qoh,Goryachev:2014nna,Berlin:2021txa}.

High-frequency gravitational-waves (HFGWs), typically above several kilohertz, are of particular interest because known astrophysical sources are unlikely to produce detectable signals in this band \cite{Aggarwal:2020olq}. Their discovery would therefore point to exotic compact objects or cosmological sources, including boson stars and dark matter stars \cite{Aggarwal:2020olq,Liebling:2012fv,Cardoso:2019rvt,Palenzuela:2017kcg,Giudice:2016zpa,Banks:2023eym}, primordial black holes \cite{Wang:2019kaf,Anantua:2008am,Dolgov:2011cq,Dong:2015yjs}, inflationary scenarios \cite{Bartolo:2016ami,Athron:2023mer,Easther:2006vd,Garcia-Bellido:2007nns,Garcia-Bellido:2007fiu,Dufaux:2007pt}, cosmological phase transitions \cite{Hasegawa:2019amx,Croon:2024mde,Kamionkowski:1993fg,Hindmarsh:2013xza,Ai:2025fqw,Guo:2020grp}, or cosmic strings \cite{Sun:2020gem,Lozanov:2019ylm,Servant:2023tua}. Detecting HFGWs is particularly challenging due to their extremely weak coupling to matter and electromagnetic fields. Although pioneering experiments such as the Holometer \cite{Holometer:2016qoh}, bulk acoustic wave (BAW) devices \cite{Goryachev:2014nna}, and OSQAR/CAST \cite{Ejlli:2019bqj} have achieved remarkable sensitivity in their respective bands, these efforts leave a vast unexplored parameter space in the GHz--THz regime.

One promising approach is the inverse Gertsenshtein effect \cite{Gertsenshtein1961}, in which a gravitational wave propagating through a magnetic field can be converted into electromagnetic radiation \cite{Gertsenshtein:1962kfm,Boccaletti1970,Zeldovich1974,Raffelt:1987im}. However, implementing this mechanism in astrophysical environments, such as neutron-star magnetospheres \cite{Li:2025eoo,Liu:2023mll} and intergalactic magnetic fields \cite{Pshirkov:2009sf,Domcke:2020yzq}, suffers from significant uncertainties in both field strength and spatial distribution. Meanwhile, terrestrial experiments relying on laboratory-scale magnetic fields \cite{Holometer:2016qoh,Goryachev:2014nna,Ejlli:2019bqj} are limited by small interaction volumes, rendering them ineffective for probing HFGWs.

The Earth's magnetic field uniquely enables inverse Gertsenshtein studies by combining laboratory and astrophysical advantages. The geomagnetic field spans planetary scales, providing an enormous region for gravitational-wave to electromagnetic-wave conversion, while simultaneously being precisely mapped and well-characterized. Leveraging these dual advantages, we analyze by far the most sensitive radio observations from the Very Large Array (VLA) and the Atacama Large Millimeter/submillimeter Array (ALMA) to search for HFGWs in the 1~GHz--1~THz range. Our results establish new constraints on the high-frequency gravitational-wave background, surpassing previous limits by up to three orders of magnitude in the relevant frequency band, which represents a significant step toward detecting exotic compact objects and cosmological phenomena.

{\it Conversion Probability.}~~The conversion between gravitational waves and electromagnetic waves originates from the coupling of metric perturbations and the electromagnetic tensor. For a gravitational wave propagating along the line of sight, the probability of conversion to electromagnetic radiation via the inverse Gertsenshtein effect is given by
\begin{equation}
    P(\Omega)=4\pi G\left|\int^{L_{\rm max}}_{0} dr \,B_{t}(r,\Omega)\right|^{2} \;,
    \label{eqn:probability}
\end{equation}
where $\Omega \equiv (\theta,\phi)$ specifies the observational direction in terms of the zenith angle $\theta$ and azimuthal angle $\phi$, $B_{t}(r,\Omega)$ denotes the tangential component of the geomagnetic field perpendicular to the propagation direction, and $r$ is the radial distance from the observer. The integral $\int B_t\,dr$ is dominated by the near-Earth region because the geomagnetic field follows an approximate dipole profile $B(r) \approx B_0 (r_\oplus/r)^3$, where $B_0 \sim 30\text{--}60\,\mu\text{T}$ at Earth's surface and $r_\oplus = 6371\,\text{km}$ is Earth's radius. For a typical line of sight extending to $L_{\rm max} = 10\,r_{\oplus}$, it ensures that 99\% of the conversion probability is accumulated within the integration range. A detailed derivation of Eq.~(\ref{eqn:probability}) is provided in the Supplemental Material.

The conversion probability in Eq.~(\ref{eqn:probability}) is determined by the line-of-sight integral of the tangential geomagnetic field component $B_{t}(r,\Omega)$.  Physically, regions with stronger transverse magnetic fields enhance the conversion efficiency, while the dipole-like structure of Earth's magnetic field creates an anisotropic conversion probability across the sky. Therefore, an accurate description of the spatial and temporal evolution of the Earth's magnetic field is essential for computing $P(\Omega)$. To this end, we adopt the International Geomagnetic Reference Field model (IGRF-14)~\cite{alken_international_2021}. As the geomagnetic field drifts slowly over time due to fluid motions in Earth's outer core, this model provides the spherical harmonic coefficients of the geomagnetic field vector at five-year intervals from 1900 to 2025, along with a forecast for the subsequent five years, thus encompassing the entire observational period of our radio telescope measurements. The IGRF model is defined in the Earth-Centered, Earth-Fixed (ECEF) coordinate system. To facilitate the line-of-sight integration of $B_{t}$, we transform the geomagnetic field from the IGRF model into the local Cartesian coordinate system centered on the telescope array, namely the East-North-Up (ENU) coordinate system (see details in the Supplemental Material).

\begin{figure}[bht]
    \centering
    \includegraphics[width=1\linewidth]{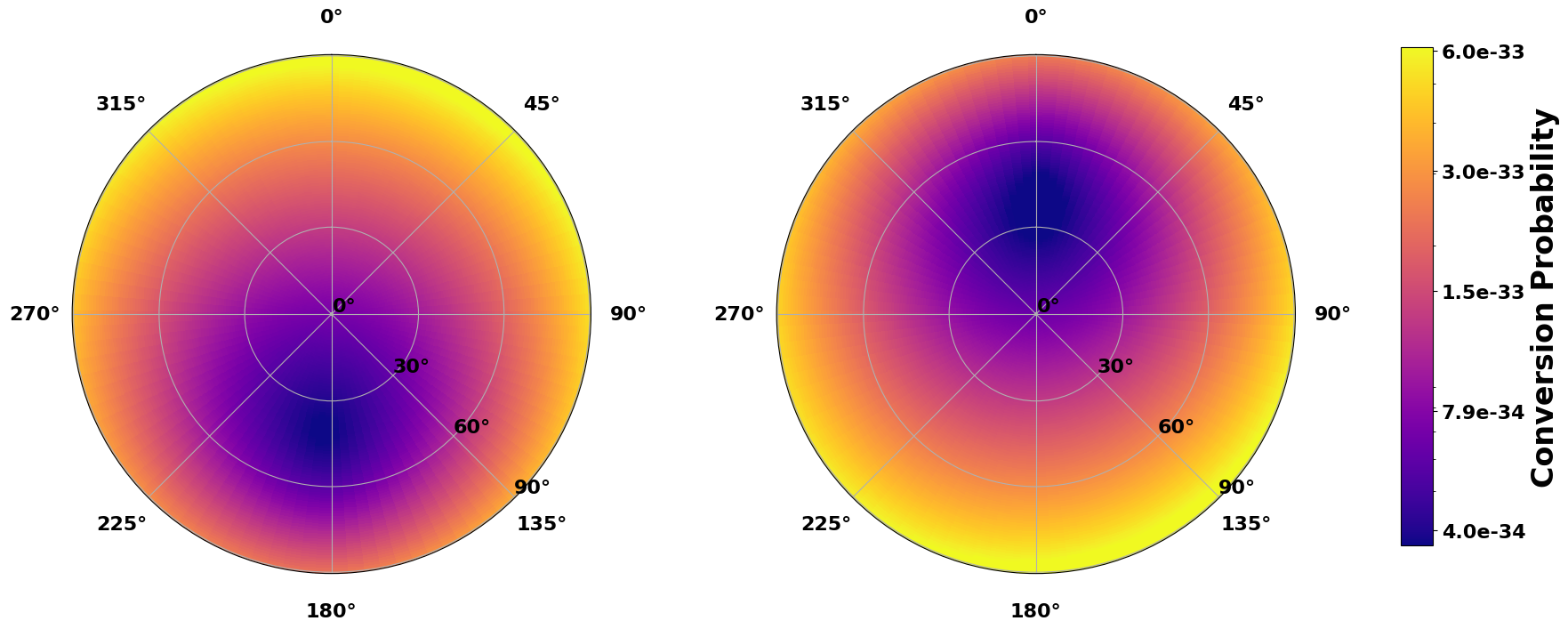}
    \caption{All-sky maps of the gravitational-wave conversion probability are shown for the VLA (left) and ALMA (right), plotted in the local East-North-Up (ENU) coordinate system. The radial coordinate corresponds to the zenith angle $\theta$, varying from $0^\circ$ (zenith) to $90^\circ$ (horizon), while the azimuth spans $0^\circ$ to $360^\circ$. The color coding represents the conversion probability, clearly distinguishing regions where the line of sight is parallel to the geomagnetic field (low probability) from those where it is perpendicular (high probability).}
    \label{fig:all_sky_map}
\end{figure}

By combining Eq.~(\ref{eqn:probability}) with the IGRF model, we present all-sky maps of the gravitational-wave-to-photon conversion probability for the VLA and ALMA with their geographic locations in Fig.~\ref{fig:all_sky_map}. The conversion probability vanishes when the gravitational-wave propagation direction aligns with the local magnetic field ($B_t = 0$), creating distinct minima in the all-sky maps. The orientation of these minima is determined by the geomagnetic field lines, which emerge from the southern magnetic pole and converge at the northern magnetic pole. Consequently, for VLA in the northern hemisphere, the magnetic field points downward and northward, so the minimum occurs when looking toward the southern celestial hemisphere where the field lines are approximately parallel to the line of sight. Conversely, for ALMA in the southern hemisphere, the field points upward and northward, producing a minimum in the northern celestial hemisphere. The typical conversion probabilities range from $P \sim 10^{-34}$ in the minima to $P \sim 10^{-33}$ in the maxima. While these values appear extremely small, they become astrophysically relevant when considering the cumulative effect of a high-frequency gravitational-wave.


{\it Upper limits on HFGWs.}~~In order to search for electromagnetic signals converted from high-frequency gravitational waves, we synergize observations from the VLA and the ALMA. This combined approach achieves ultra-wide-band frequency coverage spanning 1--1000~GHz, with the VLA covering 1--50~GHz and ALMA covering 35--1000~GHz. Observations were selected based on relatively long integration times and wide fields of view (FoV) to maximize sensitivity to a stochastic background. In total, we analyze 185 archival observations, comprising 17 from the VLA and 180 from ALMA. Data reduction was performed using the Common Astronomy Software Applications package ({\tt CASA}, v6.4.1.12), applying standard procedures for radio frequency interference (RFI) mitigation and calibration. To isolate potential diffuse conversion signals, we masked compact and extended astrophysical sources, yielding clean residual maps.

\begin{figure}[ht]
    \centering
    \includegraphics[width=0.9 \linewidth]{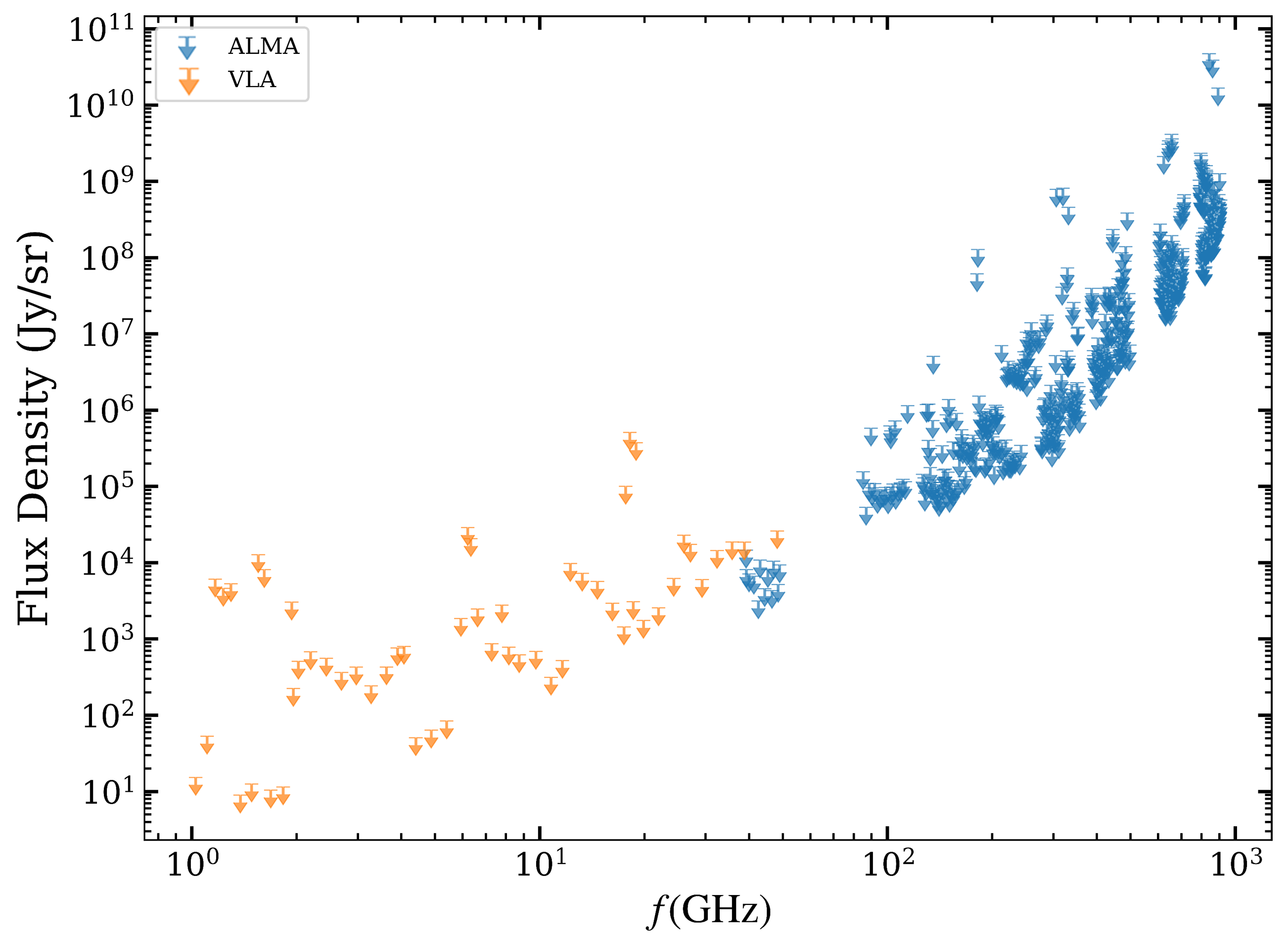}
    \caption{Upper limits on the flux density per unit solid angle at the 95\% confidence level, derived from the VLA observations (orange) spanning 1--50 GHz and the ALMA observations (blue) spanning 35--1000 GHz. These limits represent the maximum allowable flux densities from stochastic HFGW–induced electromagnetic emission.}
    \label{fig:data}
\end{figure}

In our analysis, we assume the HFGW is isotropic and unpolarized, and thus the converted electromagnetic waves inherit the spectral and angular characteristics of the gravitational-wave field, modulated by the local geomagnetic field geometry. Consequently, the flux density $I_{\gamma}(\nu)$ of the signal is directly related to the energy density spectrum of the gravitational-wave background $\Omega_{\rm GW}(\nu)$. Since no significant residual diffuse emission is detected above the instrumental noise floor, we derive upper limits on the flux densities $I_\gamma$ of the radio signals in Fig.~\ref{fig:data}. Depending on the observational conditions of target sources, e.g., zenith angles, exposure time, foreground and background emissions, the upper limits vary from one to another observation. Here we choose the lowest results in our sample to be shown. A notable gap in coverage exists between 50 and 90~GHz. This is attributed to two primary factors: (i) the severe atmospheric absorption by molecular oxygen, which peaks around $\sim 60$ GHz, and (ii) the instrumental limitations of the ALMA's Band 2 (67--90~GHz) which was not fully operational or optimized during the epochs of the selected observations. The complete analysis pipeline is detailed in the Supplement Material. All processed data products and the observational log used in this work have been publicly released~\href{https://github.com/Hong-liangTian/Supplementary-Data-File}{\faGithub}. The conversion may be affected by the ionosphere in the lower frequency range, such as below 10 GHz (see Fig.~\ref{fig:survival} in Supplement Material), because the phase mismatch induced by the plasma frequency partially cancels the conversion amplitude at specific frequencies. But the plasma effect becomes negligible above 10 GHz.


To convert these constraints into the limits on the gravitational-wave characteristic strain $h_c$, we relate the flux density of the converted photons, $I_\gamma$, to the energy density spectrum of the gravitational-wave via $I_{\gamma} = I_{\text{GW}} \cdot P$, where $I_{\text{GW}} = 2 \pi d^2\rho_{\text{GW}}/ (d \omega d \Omega) = \hbar \omega h_{c}^{2}/(2\kappa^{2})$. Since a radio telescope observes a fixed sky patch over a finite duration, the Earth's rotation sweeps this patch along a curved trajectory in the ENU coordinate frame. Furthermore, as discussed in the context of ionospheric effects, the coherence of this conversion is frequency-dependent. We therefore compute the average conversion probability $\bar{P}$ over both the observation duration $T$ and the telescope's field of view, effectively integrating along the pointing trajectory while accounting for decoherence. In practice, we discretize the observation track into 50 interpolated time points $t_i$ and approximate the integral by an arithmetic mean, $\bar{P} \approx \frac{1}{N}\sum_{i=1}^{N} P(\theta_i, \phi_i)$, where $N=50$. Using this average probability, we arrive at the final expression for the expected electromagnetic flux per steradian as,
\begin{equation}
    I_{\gamma} = \frac{\pi f h_{c}^{2}}{\kappa^{2}} \bar{P},
    \label{eqn:flux_estimate}
\end{equation}
where $\omega = 2\pi f$ is the angular frequency, and $\kappa = \sqrt{16\pi G}$ is Einstein's gravitational constant.

\begin{figure}[htp]
    \centering
    \includegraphics[width=0.9 \linewidth]{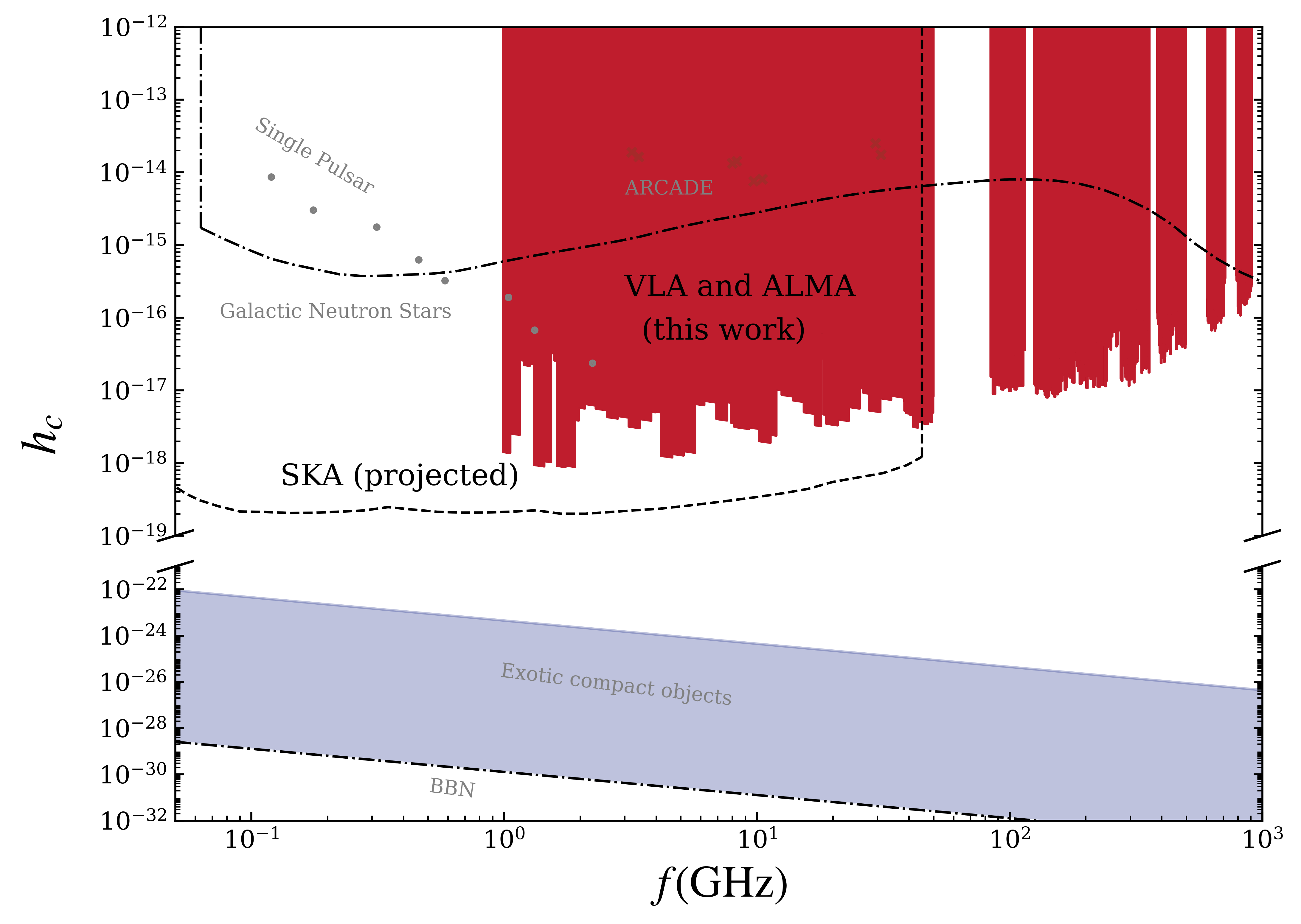}
    \caption{95\% C.L. upper limits on the characteristic strain $h_{c}$ derived from VLA and ALMA observations are displayed as red shaded regions. Other constraints from observations of galactic neutron stars~\cite{Dandoy:2024oqg} (black dash-dotted curve fit to the discrete data points reported in Ref.~\cite{Fixsen:2009xn}) and the Crab pulsar\cite{Ito:2023fcr} (gray dots) are shown for comparison. We also show limits from ARCADE from observations of the cosmic microwave background\cite{Domcke:2020yzq} (the eight frequency points are marked by crosses; the solid lines connecting them serve only to guide the eye). The limit imposed by the Big Bang Nucleosynthesis (BBN) \cite{Cyburt:2015mya} was given by the black dash-dotted line. The potential signal predictions from some exotic compact objects are shown by the blue shaded region. The projected limits of future radio observations by SKA \cite{Braun:2019gdo} are shown as dashed lines for 1 hour.}
    \label{fig:hc_sensitivity}
\end{figure}

Finally, we present the 95\% confidence level upper limits on the characteristic strain $h_c$ over the 1 -- 1000 GHz band in Fig.~\ref{fig:hc_sensitivity}. As a comparison, we also show the bounds derived from the photon fluxes induced by the magnetosphere of the galactic neutron stars~\cite{Dandoy:2024oqg}, the Crab pulsar~\cite{Ito:2023fcr}, and the cosmic magnetic field~\cite{Domcke:2020yzq}. Note that the origin of intergalactic cosmic magnetic fields remains unknown, leading to huge uncertainties in their constraints. Neutron star magnetic fields are inferred indirectly with an uncertainty of one order of magnitude or larger~\cite{Guver:2011dy}. In addition, the frequency coverage of those observations is typically not complete, but just some discrete, narrow frequency bands. Our limits are stronger than the galactic neutron star bounds by up to about 3 orders of magnitude in the 1 -- 50 GHz band. Between 50 and 1000 GHz, where ALMA provides the primary sensitivity, our results are still much more stringent than previous limits. Very importantly, the frequency coverage of this work is almost continuous, which offers the most comprehensive and robust bounds on the strain of HFGWs. The gap between 50 and 90 GHz stems from atmospheric absorption and instrumental band gaps as explained before. While current limits remain above the big bang nucleosynthesis bound~\cite{Cyburt:2015mya} for cosmological sources, they now are accessing to the parameter space of models of exotic compact objects, such as boson stars and dark matter stars~\cite{Aggarwal:2020olq,Liebling:2012fv,Cardoso:2019rvt,Palenzuela:2017kcg,Giudice:2016zpa,Banks:2023eym}. Moreover, the next-generation radio interferometer like SKA~\cite{Braun:2019gdo} is expected to increase the sensitivity significantly than current experiments. It will not only tighten our existing constraints by an additional about 1 orders of magnitude but also extend the accessible frequency window down to 50 MHz.

{\it Conclusion.}~~In summary, we have demonstrated that the Earth's magnetic field can function as a planetary-scale detector for high-frequency gravitational waves through graviton-photon conversion. By analyzing archival observations from VLA and ALMA, we performed the first HFGW search using geomagnetic conversion and obtained the strongest constraints to date across nearly continuous coverage from 1 GHz to 1 THz. Beyond the limits presented here, our work establishes a new observational strategy for exploring the high-frequency gravitational-wave Universe. Future facilities such as the SKA may substantially extend this reach and enable direct tests of a broad class of exotic compact-object and early-Universe scenarios.

{\it Acknowledgements} We would greatly thank Yuanlin Gong and Huaike Guo for discussion.
L.W. is supported by the National Natural Science Foundation of China (NNSFC) under grant No. 12335005. Q.Y. is supported by the NNSFC under grant No. 12220101003 and the Project for Young Scientists in Basic Research of Chinese Academy of Sciences under grant No. YSBR-061.

\renewcommand{\refname}{References}

\bibliography{reference}

\begin{thebibliography}{73}
\expandafter\ifx\csname natexlab\endcsname\relax\def\natexlab#1{#1}\fi
\expandafter\ifx\csname bibnamefont\endcsname\relax
  \def\bibnamefont#1{#1}\fi
\expandafter\ifx\csname bibfnamefont\endcsname\relax
  \def\bibfnamefont#1{#1}\fi
\expandafter\ifx\csname citenamefont\endcsname\relax
  \def\citenamefont#1{#1}\fi
\expandafter\ifx\csname url\endcsname\relax
  \def\url#1{\texttt{#1}}\fi
\expandafter\ifx\csname urlprefix\endcsname\relax\def\urlprefix{URL }\fi
\providecommand{\bibinfo}[2]{#2}
\providecommand{\eprint}[2][]{\url{#2}}

\bibitem[{\citenamefont{Abbott et~al.}(2016{\natexlab{a}})}]{LIGOScientific:2016aoc}
\bibinfo{author}{\bibfnamefont{B.~P.} \bibnamefont{Abbott}} \bibnamefont{et~al.} (\bibinfo{collaboration}{LIGO Scientific, Virgo}), \bibinfo{journal}{Phys. Rev. Lett.} \textbf{\bibinfo{volume}{116}}, \bibinfo{pages}{061102} (\bibinfo{year}{2016}{\natexlab{a}}), \eprint{1602.03837}.

\bibitem[{\citenamefont{Abbott et~al.}(2016{\natexlab{b}})}]{LIGOScientific:2016sjg}
\bibinfo{author}{\bibfnamefont{B.~P.} \bibnamefont{Abbott}} \bibnamefont{et~al.} (\bibinfo{collaboration}{LIGO Scientific, Virgo}), \bibinfo{journal}{Phys. Rev. Lett.} \textbf{\bibinfo{volume}{116}}, \bibinfo{pages}{241103} (\bibinfo{year}{2016}{\natexlab{b}}), \eprint{1606.04855}.

\bibitem[{\citenamefont{Abbott et~al.}(2017{\natexlab{a}})}]{LIGOScientific:2017vwq}
\bibinfo{author}{\bibfnamefont{B.~P.} \bibnamefont{Abbott}} \bibnamefont{et~al.} (\bibinfo{collaboration}{LIGO Scientific, Virgo}), \bibinfo{journal}{Phys. Rev. Lett.} \textbf{\bibinfo{volume}{119}}, \bibinfo{pages}{161101} (\bibinfo{year}{2017}{\natexlab{a}}), \eprint{1710.05832}.

\bibitem[{\citenamefont{Abbott et~al.}(2017{\natexlab{b}})}]{LIGOScientific:2017zic}
\bibinfo{author}{\bibfnamefont{B.~P.} \bibnamefont{Abbott}} \bibnamefont{et~al.} (\bibinfo{collaboration}{LIGO Scientific, Virgo, Fermi-GBM, INTEGRAL}), \bibinfo{journal}{Astrophys. J. Lett.} \textbf{\bibinfo{volume}{848}}, \bibinfo{pages}{L13} (\bibinfo{year}{2017}{\natexlab{b}}), \eprint{1710.05834}.

\bibitem[{\citenamefont{Abbott et~al.}(2019)}]{LIGOScientific:2018mvr}
\bibinfo{author}{\bibfnamefont{B.~P.} \bibnamefont{Abbott}} \bibnamefont{et~al.} (\bibinfo{collaboration}{LIGO Scientific, Virgo}), \bibinfo{journal}{Phys. Rev. X} \textbf{\bibinfo{volume}{9}}, \bibinfo{pages}{031040} (\bibinfo{year}{2019}), \eprint{1811.12907}.

\bibitem[{\citenamefont{Abbott et~al.}(2021)}]{LIGOScientific:2020ibl}
\bibinfo{author}{\bibfnamefont{R.}~\bibnamefont{Abbott}} \bibnamefont{et~al.} (\bibinfo{collaboration}{LIGO Scientific, Virgo}), \bibinfo{journal}{Phys. Rev. X} \textbf{\bibinfo{volume}{11}}, \bibinfo{pages}{021053} (\bibinfo{year}{2021}), \eprint{2010.14527}.

\bibitem[{\citenamefont{Abbott et~al.}(2023)}]{KAGRA:2021vkt}
\bibinfo{author}{\bibfnamefont{R.}~\bibnamefont{Abbott}} \bibnamefont{et~al.} (\bibinfo{collaboration}{KAGRA, VIRGO, LIGO Scientific}), \bibinfo{journal}{Phys. Rev. X} \textbf{\bibinfo{volume}{13}}, \bibinfo{pages}{041039} (\bibinfo{year}{2023}), \eprint{2111.03606}.

\bibitem[{\citenamefont{Hobbs et~al.}(2010)}]{Hobbs:2009yy}
\bibinfo{author}{\bibfnamefont{G.}~\bibnamefont{Hobbs}} \bibnamefont{et~al.}, \bibinfo{journal}{Class. Quant. Grav.} \textbf{\bibinfo{volume}{27}}, \bibinfo{pages}{084013} (\bibinfo{year}{2010}), \eprint{0911.5206}.

\bibitem[{\citenamefont{Manchester et~al.}(2013)}]{Manchester:2012za}
\bibinfo{author}{\bibfnamefont{R.~N.} \bibnamefont{Manchester}} \bibnamefont{et~al.}, \bibinfo{journal}{Publ. Astron. Soc. Austral.} \textbf{\bibinfo{volume}{30}}, \bibinfo{pages}{17} (\bibinfo{year}{2013}), \eprint{1210.6130}.

\bibitem[{\citenamefont{McLaughlin}(2013)}]{McLaughlin:2013ira}
\bibinfo{author}{\bibfnamefont{M.~A.} \bibnamefont{McLaughlin}}, \bibinfo{journal}{Class. Quant. Grav.} \textbf{\bibinfo{volume}{30}}, \bibinfo{pages}{224008} (\bibinfo{year}{2013}), \eprint{1310.0758}.

\bibitem[{\citenamefont{Janssen et~al.}(2015)}]{Janssen:2014dka}
\bibinfo{author}{\bibfnamefont{G.}~\bibnamefont{Janssen}} \bibnamefont{et~al.}, \bibinfo{journal}{PoS} \textbf{\bibinfo{volume}{AASKA14}}, \bibinfo{pages}{037} (\bibinfo{year}{2015}), \eprint{1501.00127}.

\bibitem[{\citenamefont{Lentati et~al.}(2015)}]{EPTA:2015qep}
\bibinfo{author}{\bibfnamefont{L.}~\bibnamefont{Lentati}} \bibnamefont{et~al.} (\bibinfo{collaboration}{EPTA}), \bibinfo{journal}{Mon. Not. Roy. Astron. Soc.} \textbf{\bibinfo{volume}{453}}, \bibinfo{pages}{2576} (\bibinfo{year}{2015}), \eprint{1504.03692}.

\bibitem[{\citenamefont{Smith et~al.}(2006)\citenamefont{Smith, Pierpaoli, and Kamionkowski}}]{Smith:2006nka}
\bibinfo{author}{\bibfnamefont{T.~L.} \bibnamefont{Smith}}, \bibinfo{author}{\bibfnamefont{E.}~\bibnamefont{Pierpaoli}}, \bibnamefont{and} \bibinfo{author}{\bibfnamefont{M.}~\bibnamefont{Kamionkowski}}, \bibinfo{journal}{Phys. Rev. Lett.} \textbf{\bibinfo{volume}{97}}, \bibinfo{pages}{021301} (\bibinfo{year}{2006}), \eprint{astro-ph/0603144}.

\bibitem[{\citenamefont{Pagano et~al.}(2016)\citenamefont{Pagano, Salvati, and Melchiorri}}]{Pagano:2015hma}
\bibinfo{author}{\bibfnamefont{L.}~\bibnamefont{Pagano}}, \bibinfo{author}{\bibfnamefont{L.}~\bibnamefont{Salvati}}, \bibnamefont{and} \bibinfo{author}{\bibfnamefont{A.}~\bibnamefont{Melchiorri}}, \bibinfo{journal}{Phys. Lett. B} \textbf{\bibinfo{volume}{760}}, \bibinfo{pages}{823} (\bibinfo{year}{2016}), \eprint{1508.02393}.

\bibitem[{\citenamefont{Clarke et~al.}(2020)\citenamefont{Clarke, Copeland, and Moss}}]{Clarke:2020bil}
\bibinfo{author}{\bibfnamefont{T.~J.} \bibnamefont{Clarke}}, \bibinfo{author}{\bibfnamefont{E.~J.} \bibnamefont{Copeland}}, \bibnamefont{and} \bibinfo{author}{\bibfnamefont{A.}~\bibnamefont{Moss}}, \bibinfo{journal}{JCAP} \textbf{\bibinfo{volume}{10}}, \bibinfo{pages}{002} (\bibinfo{year}{2020}), \eprint{2004.11396}.

\bibitem[{\citenamefont{Agazie et~al.}(2023)}]{NANOGrav:2023gor}
\bibinfo{author}{\bibfnamefont{G.}~\bibnamefont{Agazie}} \bibnamefont{et~al.} (\bibinfo{collaboration}{NANOGrav}), \bibinfo{journal}{Astrophys. J. Lett.} \textbf{\bibinfo{volume}{951}}, \bibinfo{pages}{L8} (\bibinfo{year}{2023}), \eprint{2306.16213}.

\bibitem[{\citenamefont{Xu et~al.}(2023)}]{Xu:2023wog}
\bibinfo{author}{\bibfnamefont{H.}~\bibnamefont{Xu}} \bibnamefont{et~al.}, \bibinfo{journal}{Res. Astron. Astrophys.} \textbf{\bibinfo{volume}{23}}, \bibinfo{pages}{075024} (\bibinfo{year}{2023}), \eprint{2306.16216}.

\bibitem[{\citenamefont{Reardon et~al.}(2023)}]{Reardon:2023gzh}
\bibinfo{author}{\bibfnamefont{D.~J.} \bibnamefont{Reardon}} \bibnamefont{et~al.}, \bibinfo{journal}{Astrophys. J. Lett.} \textbf{\bibinfo{volume}{951}}, \bibinfo{pages}{L6} (\bibinfo{year}{2023}), \eprint{2306.16215}.

\bibitem[{\citenamefont{Antoniadis et~al.}(2023)}]{EPTA:2023fyk}
\bibinfo{author}{\bibfnamefont{J.}~\bibnamefont{Antoniadis}} \bibnamefont{et~al.} (\bibinfo{collaboration}{EPTA, InPTA:}), \bibinfo{journal}{Astron. Astrophys.} \textbf{\bibinfo{volume}{678}}, \bibinfo{pages}{A50} (\bibinfo{year}{2023}), \eprint{2306.16214}.

\bibitem[{\citenamefont{Miles et~al.}(2024)}]{Miles:2024seg}
\bibinfo{author}{\bibfnamefont{M.~T.} \bibnamefont{Miles}} \bibnamefont{et~al.}, \bibinfo{journal}{Mon. Not. Roy. Astron. Soc.} \textbf{\bibinfo{volume}{536}}, \bibinfo{pages}{1489} (\bibinfo{year}{2024}), \eprint{2412.01153}.

\bibitem[{\citenamefont{Domcke et~al.}(2022)\citenamefont{Domcke, Garcia-Cely, and Rodd}}]{Domcke:2022rgu}
\bibinfo{author}{\bibfnamefont{V.}~\bibnamefont{Domcke}}, \bibinfo{author}{\bibfnamefont{C.}~\bibnamefont{Garcia-Cely}}, \bibnamefont{and} \bibinfo{author}{\bibfnamefont{N.~L.} \bibnamefont{Rodd}}, \bibinfo{journal}{Phys. Rev. Lett.} \textbf{\bibinfo{volume}{129}}, \bibinfo{pages}{041101} (\bibinfo{year}{2022}), \eprint{2202.00695}.

\bibitem[{\citenamefont{Ejlli et~al.}(2019)\citenamefont{Ejlli, Ejlli, Cruise, Pisano, and Grote}}]{Ejlli:2019bqj}
\bibinfo{author}{\bibfnamefont{A.}~\bibnamefont{Ejlli}}, \bibinfo{author}{\bibfnamefont{D.}~\bibnamefont{Ejlli}}, \bibinfo{author}{\bibfnamefont{A.~M.} \bibnamefont{Cruise}}, \bibinfo{author}{\bibfnamefont{G.}~\bibnamefont{Pisano}}, \bibnamefont{and} \bibinfo{author}{\bibfnamefont{H.}~\bibnamefont{Grote}}, \bibinfo{journal}{Eur. Phys. J. C} \textbf{\bibinfo{volume}{79}}, \bibinfo{pages}{1032} (\bibinfo{year}{2019}), \eprint{1908.00232}.

\bibitem[{\citenamefont{Chou et~al.}(2017)}]{Holometer:2016qoh}
\bibinfo{author}{\bibfnamefont{A.~S.} \bibnamefont{Chou}} \bibnamefont{et~al.} (\bibinfo{collaboration}{Holometer}), \bibinfo{journal}{Phys. Rev. D} \textbf{\bibinfo{volume}{95}}, \bibinfo{pages}{063002} (\bibinfo{year}{2017}), \eprint{1611.05560}.

\bibitem[{\citenamefont{Goryachev et~al.}(2014)\citenamefont{Goryachev, Ivanov, van Kann, Galliou, and Tobar}}]{Goryachev:2014nna}
\bibinfo{author}{\bibfnamefont{M.}~\bibnamefont{Goryachev}}, \bibinfo{author}{\bibfnamefont{E.~N.} \bibnamefont{Ivanov}}, \bibinfo{author}{\bibfnamefont{F.}~\bibnamefont{van Kann}}, \bibinfo{author}{\bibfnamefont{S.}~\bibnamefont{Galliou}}, \bibnamefont{and} \bibinfo{author}{\bibfnamefont{M.~E.} \bibnamefont{Tobar}}, \bibinfo{journal}{Appl. Phys. Lett.} \textbf{\bibinfo{volume}{105}}, \bibinfo{pages}{153505} (\bibinfo{year}{2014}), \eprint{1410.4293}.

\bibitem[{\citenamefont{Berlin et~al.}(2022)\citenamefont{Berlin, Blas, Tito~D'Agnolo, Ellis, Harnik, Kahn, and Sch{\"u}tte-Engel}}]{Berlin:2021txa}
\bibinfo{author}{\bibfnamefont{A.}~\bibnamefont{Berlin}}, \bibinfo{author}{\bibfnamefont{D.}~\bibnamefont{Blas}}, \bibinfo{author}{\bibfnamefont{R.}~\bibnamefont{Tito~D'Agnolo}}, \bibinfo{author}{\bibfnamefont{S.~A.~R.} \bibnamefont{Ellis}}, \bibinfo{author}{\bibfnamefont{R.}~\bibnamefont{Harnik}}, \bibinfo{author}{\bibfnamefont{Y.}~\bibnamefont{Kahn}}, \bibnamefont{and} \bibinfo{author}{\bibfnamefont{J.}~\bibnamefont{Sch{\"u}tte-Engel}}, \bibinfo{journal}{Phys. Rev. D} \textbf{\bibinfo{volume}{105}}, \bibinfo{pages}{116011} (\bibinfo{year}{2022}), \eprint{2112.11465}.

\bibitem[{\citenamefont{Aggarwal et~al.}(2021)}]{Aggarwal:2020olq}
\bibinfo{author}{\bibfnamefont{N.}~\bibnamefont{Aggarwal}} \bibnamefont{et~al.}, \bibinfo{journal}{Living Rev. Rel.} \textbf{\bibinfo{volume}{24}}, \bibinfo{pages}{4} (\bibinfo{year}{2021}), \eprint{2011.12414}.

\bibitem[{\citenamefont{Liebling and Palenzuela}(2023)}]{Liebling:2012fv}
\bibinfo{author}{\bibfnamefont{S.~L.} \bibnamefont{Liebling}} \bibnamefont{and} \bibinfo{author}{\bibfnamefont{C.}~\bibnamefont{Palenzuela}}, \bibinfo{journal}{Living Rev. Rel.} \textbf{\bibinfo{volume}{26}}, \bibinfo{pages}{1} (\bibinfo{year}{2023}), \eprint{1202.5809}.

\bibitem[{\citenamefont{Cardoso and Pani}(2019)}]{Cardoso:2019rvt}
\bibinfo{author}{\bibfnamefont{V.}~\bibnamefont{Cardoso}} \bibnamefont{and} \bibinfo{author}{\bibfnamefont{P.}~\bibnamefont{Pani}}, \bibinfo{journal}{Living Rev. Rel.} \textbf{\bibinfo{volume}{22}}, \bibinfo{pages}{4} (\bibinfo{year}{2019}), \eprint{1904.05363}.

\bibitem[{\citenamefont{Palenzuela et~al.}(2017)\citenamefont{Palenzuela, Pani, Bezares, Cardoso, Lehner, and Liebling}}]{Palenzuela:2017kcg}
\bibinfo{author}{\bibfnamefont{C.}~\bibnamefont{Palenzuela}}, \bibinfo{author}{\bibfnamefont{P.}~\bibnamefont{Pani}}, \bibinfo{author}{\bibfnamefont{M.}~\bibnamefont{Bezares}}, \bibinfo{author}{\bibfnamefont{V.}~\bibnamefont{Cardoso}}, \bibinfo{author}{\bibfnamefont{L.}~\bibnamefont{Lehner}}, \bibnamefont{and} \bibinfo{author}{\bibfnamefont{S.}~\bibnamefont{Liebling}}, \bibinfo{journal}{Phys. Rev. D} \textbf{\bibinfo{volume}{96}}, \bibinfo{pages}{104058} (\bibinfo{year}{2017}), \eprint{1710.09432}.

\bibitem[{\citenamefont{Giudice et~al.}(2016)\citenamefont{Giudice, McCullough, and Urbano}}]{Giudice:2016zpa}
\bibinfo{author}{\bibfnamefont{G.~F.} \bibnamefont{Giudice}}, \bibinfo{author}{\bibfnamefont{M.}~\bibnamefont{McCullough}}, \bibnamefont{and} \bibinfo{author}{\bibfnamefont{A.}~\bibnamefont{Urbano}}, \bibinfo{journal}{JCAP} \textbf{\bibinfo{volume}{10}}, \bibinfo{pages}{001} (\bibinfo{year}{2016}), \eprint{1605.01209}.

\bibitem[{\citenamefont{Banks et~al.}(2023)\citenamefont{Banks, Grabowska, and McCullough}}]{Banks:2023eym}
\bibinfo{author}{\bibfnamefont{H.}~\bibnamefont{Banks}}, \bibinfo{author}{\bibfnamefont{D.~M.} \bibnamefont{Grabowska}}, \bibnamefont{and} \bibinfo{author}{\bibfnamefont{M.}~\bibnamefont{McCullough}}, \bibinfo{journal}{Phys. Rev. D} \textbf{\bibinfo{volume}{108}}, \bibinfo{pages}{035017} (\bibinfo{year}{2023}), \eprint{2302.07887}.

\bibitem[{\citenamefont{Wang et~al.}(2019)\citenamefont{Wang, Terada, and Kohri}}]{Wang:2019kaf}
\bibinfo{author}{\bibfnamefont{S.}~\bibnamefont{Wang}}, \bibinfo{author}{\bibfnamefont{T.}~\bibnamefont{Terada}}, \bibnamefont{and} \bibinfo{author}{\bibfnamefont{K.}~\bibnamefont{Kohri}}, \bibinfo{journal}{Phys. Rev. D} \textbf{\bibinfo{volume}{99}}, \bibinfo{pages}{103531} (\bibinfo{year}{2019}), \bibinfo{note}{[Erratum: Phys.Rev.D 101, 069901 (2020)]}, \eprint{1903.05924}.

\bibitem[{\citenamefont{Anantua et~al.}(2009)\citenamefont{Anantua, Easther, and Giblin}}]{Anantua:2008am}
\bibinfo{author}{\bibfnamefont{R.}~\bibnamefont{Anantua}}, \bibinfo{author}{\bibfnamefont{R.}~\bibnamefont{Easther}}, \bibnamefont{and} \bibinfo{author}{\bibfnamefont{J.~T.} \bibnamefont{Giblin}}, \bibinfo{journal}{Phys. Rev. Lett.} \textbf{\bibinfo{volume}{103}}, \bibinfo{pages}{111303} (\bibinfo{year}{2009}), \eprint{0812.0825}.

\bibitem[{\citenamefont{Dolgov and Ejlli}(2011)}]{Dolgov:2011cq}
\bibinfo{author}{\bibfnamefont{A.~D.} \bibnamefont{Dolgov}} \bibnamefont{and} \bibinfo{author}{\bibfnamefont{D.}~\bibnamefont{Ejlli}}, \bibinfo{journal}{Phys. Rev. D} \textbf{\bibinfo{volume}{84}}, \bibinfo{pages}{024028} (\bibinfo{year}{2011}), \eprint{1105.2303}.

\bibitem[{\citenamefont{Dong et~al.}(2016)\citenamefont{Dong, Kinney, and Stojkovic}}]{Dong:2015yjs}
\bibinfo{author}{\bibfnamefont{R.}~\bibnamefont{Dong}}, \bibinfo{author}{\bibfnamefont{W.~H.} \bibnamefont{Kinney}}, \bibnamefont{and} \bibinfo{author}{\bibfnamefont{D.}~\bibnamefont{Stojkovic}}, \bibinfo{journal}{JCAP} \textbf{\bibinfo{volume}{10}}, \bibinfo{pages}{034} (\bibinfo{year}{2016}), \eprint{1511.05642}.

\bibitem[{\citenamefont{Bartolo et~al.}(2016)}]{Bartolo:2016ami}
\bibinfo{author}{\bibfnamefont{N.}~\bibnamefont{Bartolo}} \bibnamefont{et~al.}, \bibinfo{journal}{JCAP} \textbf{\bibinfo{volume}{12}}, \bibinfo{pages}{026} (\bibinfo{year}{2016}), \eprint{1610.06481}.

\bibitem[{\citenamefont{Athron et~al.}(2024)\citenamefont{Athron, Fowlie, Lu, Morris, Wu, Wu, and Xu}}]{Athron:2023mer}
\bibinfo{author}{\bibfnamefont{P.}~\bibnamefont{Athron}}, \bibinfo{author}{\bibfnamefont{A.}~\bibnamefont{Fowlie}}, \bibinfo{author}{\bibfnamefont{C.-T.} \bibnamefont{Lu}}, \bibinfo{author}{\bibfnamefont{L.}~\bibnamefont{Morris}}, \bibinfo{author}{\bibfnamefont{L.}~\bibnamefont{Wu}}, \bibinfo{author}{\bibfnamefont{Y.}~\bibnamefont{Wu}}, \bibnamefont{and} \bibinfo{author}{\bibfnamefont{Z.}~\bibnamefont{Xu}}, \bibinfo{journal}{Phys. Rev. Lett.} \textbf{\bibinfo{volume}{132}}, \bibinfo{pages}{221001} (\bibinfo{year}{2024}), \eprint{2306.17239}.

\bibitem[{\citenamefont{Easther et~al.}(2007)\citenamefont{Easther, Giblin, and Lim}}]{Easther:2006vd}
\bibinfo{author}{\bibfnamefont{R.}~\bibnamefont{Easther}}, \bibinfo{author}{\bibfnamefont{J.~T.} \bibnamefont{Giblin}, \bibfnamefont{Jr.}}, \bibnamefont{and} \bibinfo{author}{\bibfnamefont{E.~A.} \bibnamefont{Lim}}, \bibinfo{journal}{Phys. Rev. Lett.} \textbf{\bibinfo{volume}{99}}, \bibinfo{pages}{221301} (\bibinfo{year}{2007}), \eprint{astro-ph/0612294}.

\bibitem[{\citenamefont{Garcia-Bellido and Figueroa}(2007)}]{Garcia-Bellido:2007nns}
\bibinfo{author}{\bibfnamefont{J.}~\bibnamefont{Garcia-Bellido}} \bibnamefont{and} \bibinfo{author}{\bibfnamefont{D.~G.} \bibnamefont{Figueroa}}, \bibinfo{journal}{Phys. Rev. Lett.} \textbf{\bibinfo{volume}{98}}, \bibinfo{pages}{061302} (\bibinfo{year}{2007}), \eprint{astro-ph/0701014}.

\bibitem[{\citenamefont{Garcia-Bellido et~al.}(2008)\citenamefont{Garcia-Bellido, Figueroa, and Sastre}}]{Garcia-Bellido:2007fiu}
\bibinfo{author}{\bibfnamefont{J.}~\bibnamefont{Garcia-Bellido}}, \bibinfo{author}{\bibfnamefont{D.~G.} \bibnamefont{Figueroa}}, \bibnamefont{and} \bibinfo{author}{\bibfnamefont{A.}~\bibnamefont{Sastre}}, \bibinfo{journal}{Phys. Rev. D} \textbf{\bibinfo{volume}{77}}, \bibinfo{pages}{043517} (\bibinfo{year}{2008}), \eprint{0707.0839}.

\bibitem[{\citenamefont{Dufaux et~al.}(2007)\citenamefont{Dufaux, Bergman, Felder, Kofman, and Uzan}}]{Dufaux:2007pt}
\bibinfo{author}{\bibfnamefont{J.~F.} \bibnamefont{Dufaux}}, \bibinfo{author}{\bibfnamefont{A.}~\bibnamefont{Bergman}}, \bibinfo{author}{\bibfnamefont{G.~N.} \bibnamefont{Felder}}, \bibinfo{author}{\bibfnamefont{L.}~\bibnamefont{Kofman}}, \bibnamefont{and} \bibinfo{author}{\bibfnamefont{J.-P.} \bibnamefont{Uzan}}, \bibinfo{journal}{Phys. Rev. D} \textbf{\bibinfo{volume}{76}}, \bibinfo{pages}{123517} (\bibinfo{year}{2007}), \eprint{0707.0875}.

\bibitem[{\citenamefont{Hasegawa et~al.}(2019)\citenamefont{Hasegawa, Okada, and Seto}}]{Hasegawa:2019amx}
\bibinfo{author}{\bibfnamefont{T.}~\bibnamefont{Hasegawa}}, \bibinfo{author}{\bibfnamefont{N.}~\bibnamefont{Okada}}, \bibnamefont{and} \bibinfo{author}{\bibfnamefont{O.}~\bibnamefont{Seto}}, \bibinfo{journal}{Phys. Rev. D} \textbf{\bibinfo{volume}{99}}, \bibinfo{pages}{095039} (\bibinfo{year}{2019}), \eprint{1904.03020}.

\bibitem[{\citenamefont{Croon and Weir}(2024)}]{Croon:2024mde}
\bibinfo{author}{\bibfnamefont{D.}~\bibnamefont{Croon}} \bibnamefont{and} \bibinfo{author}{\bibfnamefont{D.~J.} \bibnamefont{Weir}}, \bibinfo{journal}{Contemp. Phys.} \textbf{\bibinfo{volume}{65}}, \bibinfo{pages}{75} (\bibinfo{year}{2024}), \eprint{2410.21509}.

\bibitem[{\citenamefont{Kamionkowski et~al.}(1994)\citenamefont{Kamionkowski, Kosowsky, and Turner}}]{Kamionkowski:1993fg}
\bibinfo{author}{\bibfnamefont{M.}~\bibnamefont{Kamionkowski}}, \bibinfo{author}{\bibfnamefont{A.}~\bibnamefont{Kosowsky}}, \bibnamefont{and} \bibinfo{author}{\bibfnamefont{M.~S.} \bibnamefont{Turner}}, \bibinfo{journal}{Phys. Rev. D} \textbf{\bibinfo{volume}{49}}, \bibinfo{pages}{2837} (\bibinfo{year}{1994}), \eprint{astro-ph/9310044}.

\bibitem[{\citenamefont{Hindmarsh et~al.}(2014)\citenamefont{Hindmarsh, Huber, Rummukainen, and Weir}}]{Hindmarsh:2013xza}
\bibinfo{author}{\bibfnamefont{M.}~\bibnamefont{Hindmarsh}}, \bibinfo{author}{\bibfnamefont{S.~J.} \bibnamefont{Huber}}, \bibinfo{author}{\bibfnamefont{K.}~\bibnamefont{Rummukainen}}, \bibnamefont{and} \bibinfo{author}{\bibfnamefont{D.~J.} \bibnamefont{Weir}}, \bibinfo{journal}{Phys. Rev. Lett.} \textbf{\bibinfo{volume}{112}}, \bibinfo{pages}{041301} (\bibinfo{year}{2014}), \eprint{1304.2433}.

\bibitem[{\citenamefont{Ai}(2026)}]{Ai:2025fqw}
\bibinfo{author}{\bibfnamefont{W.-Y.} \bibnamefont{Ai}}, \bibinfo{journal}{Phys. Rev. D} \textbf{\bibinfo{volume}{113}}, \bibinfo{pages}{056007} (\bibinfo{year}{2026}), \eprint{2508.02794}.

\bibitem[{\citenamefont{Guo et~al.}(2021)\citenamefont{Guo, Sinha, Vagie, and White}}]{Guo:2020grp}
\bibinfo{author}{\bibfnamefont{H.-K.} \bibnamefont{Guo}}, \bibinfo{author}{\bibfnamefont{K.}~\bibnamefont{Sinha}}, \bibinfo{author}{\bibfnamefont{D.}~\bibnamefont{Vagie}}, \bibnamefont{and} \bibinfo{author}{\bibfnamefont{G.}~\bibnamefont{White}}, \bibinfo{journal}{JCAP} \textbf{\bibinfo{volume}{01}}, \bibinfo{pages}{001} (\bibinfo{year}{2021}), \eprint{2007.08537}.

\bibitem[{\citenamefont{Sun and Zhang}(2021)}]{Sun:2020gem}
\bibinfo{author}{\bibfnamefont{S.}~\bibnamefont{Sun}} \bibnamefont{and} \bibinfo{author}{\bibfnamefont{Y.-L.} \bibnamefont{Zhang}}, \bibinfo{journal}{Phys. Rev. D} \textbf{\bibinfo{volume}{104}}, \bibinfo{pages}{103009} (\bibinfo{year}{2021}), \eprint{2003.10527}.

\bibitem[{\citenamefont{Lozanov and Amin}(2019)}]{Lozanov:2019ylm}
\bibinfo{author}{\bibfnamefont{K.~D.} \bibnamefont{Lozanov}} \bibnamefont{and} \bibinfo{author}{\bibfnamefont{M.~A.} \bibnamefont{Amin}}, \bibinfo{journal}{Phys. Rev. D} \textbf{\bibinfo{volume}{99}}, \bibinfo{pages}{123504} (\bibinfo{year}{2019}), \eprint{1902.06736}.

\bibitem[{\citenamefont{Servant and Simakachorn}(2024)}]{Servant:2023tua}
\bibinfo{author}{\bibfnamefont{G.}~\bibnamefont{Servant}} \bibnamefont{and} \bibinfo{author}{\bibfnamefont{P.}~\bibnamefont{Simakachorn}}, \bibinfo{journal}{Phys. Rev. D} \textbf{\bibinfo{volume}{109}}, \bibinfo{pages}{103538} (\bibinfo{year}{2024}), \eprint{2312.09281}.

\bibitem[{\citenamefont{Gertsenshtein}(1961)}]{Gertsenshtein1961}
\bibinfo{author}{\bibfnamefont{M.~E.} \bibnamefont{Gertsenshtein}}, \bibinfo{journal}{Journal of Experimental and Theoretical Physics} \textbf{\bibinfo{volume}{41}}, \bibinfo{pages}{113} (\bibinfo{year}{1961}), \bibinfo{note}{english translation in \textit{Soviet Physics JETP}, Vol. 14, pp. 84--85 (1962)}.

\bibitem[{\citenamefont{Gertsenshtein and Pustovoit}(1962)}]{Gertsenshtein:1962kfm}
\bibinfo{author}{\bibfnamefont{M.~E.} \bibnamefont{Gertsenshtein}} \bibnamefont{and} \bibinfo{author}{\bibfnamefont{V.~I.} \bibnamefont{Pustovoit}}, \bibinfo{journal}{Sov. Phys. JETP} \textbf{\bibinfo{volume}{16}}, \bibinfo{pages}{433} (\bibinfo{year}{1962}).

\bibitem[{\citenamefont{Boccaletti et~al.}(1970)\citenamefont{Boccaletti, De~Sabbata, Fortini, and Gualdi}}]{Boccaletti1970}
\bibinfo{author}{\bibfnamefont{D.}~\bibnamefont{Boccaletti}}, \bibinfo{author}{\bibfnamefont{V.}~\bibnamefont{De~Sabbata}}, \bibinfo{author}{\bibfnamefont{P.}~\bibnamefont{Fortini}}, \bibnamefont{and} \bibinfo{author}{\bibfnamefont{C.}~\bibnamefont{Gualdi}}, \bibinfo{journal}{Il Nuovo Cimento B (1965-1970)} \textbf{\bibinfo{volume}{70}}, \bibinfo{pages}{129} (\bibinfo{year}{1970}).

\bibitem[{\citenamefont{Zel'dovich}(1974)}]{Zeldovich1974}
\bibinfo{author}{\bibfnamefont{Y.~B.} \bibnamefont{Zel'dovich}}, \bibinfo{journal}{Soviet Journal of Experimental and Theoretical Physics} \textbf{\bibinfo{volume}{38}}, \bibinfo{pages}{652} (\bibinfo{year}{1974}).

\bibitem[{\citenamefont{Raffelt and Stodolsky}(1988)}]{Raffelt:1987im}
\bibinfo{author}{\bibfnamefont{G.}~\bibnamefont{Raffelt}} \bibnamefont{and} \bibinfo{author}{\bibfnamefont{L.}~\bibnamefont{Stodolsky}}, \bibinfo{journal}{Phys. Rev. D} \textbf{\bibinfo{volume}{37}}, \bibinfo{pages}{1237} (\bibinfo{year}{1988}).

\bibitem[{\citenamefont{Li et~al.}(2025)\citenamefont{Li, Hong, and Zhang}}]{Li:2025eoo}
\bibinfo{author}{\bibfnamefont{J.-K.} \bibnamefont{Li}}, \bibinfo{author}{\bibfnamefont{W.}~\bibnamefont{Hong}}, \bibnamefont{and} \bibinfo{author}{\bibfnamefont{T.-J.} \bibnamefont{Zhang}}, \bibinfo{journal}{Astrophys. J.} \textbf{\bibinfo{volume}{985}}, \bibinfo{pages}{137} (\bibinfo{year}{2025}), \eprint{2504.13115}.

\bibitem[{\citenamefont{Liu et~al.}(2024)\citenamefont{Liu, Ren, and Zhang}}]{Liu:2023mll}
\bibinfo{author}{\bibfnamefont{T.}~\bibnamefont{Liu}}, \bibinfo{author}{\bibfnamefont{J.}~\bibnamefont{Ren}}, \bibnamefont{and} \bibinfo{author}{\bibfnamefont{C.}~\bibnamefont{Zhang}}, \bibinfo{journal}{Phys. Rev. Lett.} \textbf{\bibinfo{volume}{132}}, \bibinfo{pages}{131402} (\bibinfo{year}{2024}), \eprint{2305.01832}.

\bibitem[{\citenamefont{Pshirkov and Baskaran}(2009)}]{Pshirkov:2009sf}
\bibinfo{author}{\bibfnamefont{M.~S.} \bibnamefont{Pshirkov}} \bibnamefont{and} \bibinfo{author}{\bibfnamefont{D.}~\bibnamefont{Baskaran}}, \bibinfo{journal}{Phys. Rev. D} \textbf{\bibinfo{volume}{80}}, \bibinfo{pages}{042002} (\bibinfo{year}{2009}), \eprint{0903.4160}.

\bibitem[{\citenamefont{Domcke and Garcia-Cely}(2021)}]{Domcke:2020yzq}
\bibinfo{author}{\bibfnamefont{V.}~\bibnamefont{Domcke}} \bibnamefont{and} \bibinfo{author}{\bibfnamefont{C.}~\bibnamefont{Garcia-Cely}}, \bibinfo{journal}{Phys. Rev. Lett.} \textbf{\bibinfo{volume}{126}}, \bibinfo{pages}{021104} (\bibinfo{year}{2021}), \eprint{2006.01161}.

\bibitem[{\citenamefont{Alken et~al.}(2021)\citenamefont{Alken, Thébault, Beggan, Amit, Aubert, Baerenzung, Bondar, Brown, Califf, Chambodut et~al.}}]{alken_international_2021}
\bibinfo{author}{\bibfnamefont{P.}~\bibnamefont{Alken}}, \bibinfo{author}{\bibfnamefont{E.}~\bibnamefont{Thébault}}, \bibinfo{author}{\bibfnamefont{C.~D.} \bibnamefont{Beggan}}, \bibinfo{author}{\bibfnamefont{H.}~\bibnamefont{Amit}}, \bibinfo{author}{\bibfnamefont{J.}~\bibnamefont{Aubert}}, \bibinfo{author}{\bibfnamefont{J.}~\bibnamefont{Baerenzung}}, \bibinfo{author}{\bibfnamefont{T.~N.} \bibnamefont{Bondar}}, \bibinfo{author}{\bibfnamefont{W.~J.} \bibnamefont{Brown}}, \bibinfo{author}{\bibfnamefont{S.}~\bibnamefont{Califf}}, \bibinfo{author}{\bibfnamefont{A.}~\bibnamefont{Chambodut}}, \bibnamefont{et~al.}, \bibinfo{journal}{Earth, Planets and Space} \textbf{\bibinfo{volume}{73}}, \bibinfo{pages}{49} (\bibinfo{year}{2021}), ISSN \bibinfo{issn}{1880-5981}, \urlprefix\url{https://doi.org/10.1186/s40623-020-01288-x}.

\bibitem[{\citenamefont{Dandoy et~al.}(2024)\citenamefont{Dandoy, Bert{\'o}lez-Mart{\'\i}nez, and Costa}}]{Dandoy:2024oqg}
\bibinfo{author}{\bibfnamefont{V.}~\bibnamefont{Dandoy}}, \bibinfo{author}{\bibfnamefont{T.}~\bibnamefont{Bert{\'o}lez-Mart{\'\i}nez}}, \bibnamefont{and} \bibinfo{author}{\bibfnamefont{F.}~\bibnamefont{Costa}}, \bibinfo{journal}{JCAP} \textbf{\bibinfo{volume}{12}}, \bibinfo{pages}{023} (\bibinfo{year}{2024}), \eprint{2402.14092}.

\bibitem[{\citenamefont{Ito et~al.}(2024)\citenamefont{Ito, Kohri, and Nakayama}}]{Ito:2023fcr}
\bibinfo{author}{\bibfnamefont{A.}~\bibnamefont{Ito}}, \bibinfo{author}{\bibfnamefont{K.}~\bibnamefont{Kohri}}, \bibnamefont{and} \bibinfo{author}{\bibfnamefont{K.}~\bibnamefont{Nakayama}}, \bibinfo{journal}{Phys. Rev. D} \textbf{\bibinfo{volume}{109}}, \bibinfo{pages}{063026} (\bibinfo{year}{2024}), \eprint{2305.13984}.

\bibitem[{\citenamefont{Guver et~al.}(2011)\citenamefont{Guver, Gogus, and Ozel}}]{Guver:2011dy}
\bibinfo{author}{\bibfnamefont{T.}~\bibnamefont{Guver}}, \bibinfo{author}{\bibfnamefont{E.}~\bibnamefont{Gogus}}, \bibnamefont{and} \bibinfo{author}{\bibfnamefont{F.}~\bibnamefont{Ozel}}, \bibinfo{journal}{Mon. Not. Roy. Astron. Soc.} \textbf{\bibinfo{volume}{418}}, \bibinfo{pages}{2773} (\bibinfo{year}{2011}), \eprint{1103.3024}.

\bibitem[{\citenamefont{Cyburt et~al.}(2016)\citenamefont{Cyburt, Fields, Olive, and Yeh}}]{Cyburt:2015mya}
\bibinfo{author}{\bibfnamefont{R.~H.} \bibnamefont{Cyburt}}, \bibinfo{author}{\bibfnamefont{B.~D.} \bibnamefont{Fields}}, \bibinfo{author}{\bibfnamefont{K.~A.} \bibnamefont{Olive}}, \bibnamefont{and} \bibinfo{author}{\bibfnamefont{T.-H.} \bibnamefont{Yeh}}, \bibinfo{journal}{Rev. Mod. Phys.} \textbf{\bibinfo{volume}{88}}, \bibinfo{pages}{015004} (\bibinfo{year}{2016}), \eprint{1505.01076}.

\bibitem[{\citenamefont{Braun et~al.}(2019)\citenamefont{Braun, Bonaldi, Bourke, Keane, and Wagg}}]{Braun:2019gdo}
\bibinfo{author}{\bibfnamefont{R.}~\bibnamefont{Braun}}, \bibinfo{author}{\bibfnamefont{A.}~\bibnamefont{Bonaldi}}, \bibinfo{author}{\bibfnamefont{T.}~\bibnamefont{Bourke}}, \bibinfo{author}{\bibfnamefont{E.}~\bibnamefont{Keane}}, \bibnamefont{and} \bibinfo{author}{\bibfnamefont{J.}~\bibnamefont{Wagg}} (\bibinfo{year}{2019}), \eprint{1912.12699}.

\bibitem[{\citenamefont{Fixsen et~al.}(2011)}]{Fixsen:2009xn}
\bibinfo{author}{\bibfnamefont{D.~J.} \bibnamefont{Fixsen}} \bibnamefont{et~al.}, \bibinfo{journal}{Astrophys. J.} \textbf{\bibinfo{volume}{734}}, \bibinfo{pages}{5} (\bibinfo{year}{2011}), \eprint{0901.0555}.

\bibitem[{\citenamefont{Ejlli and Thandlam}(2019)}]{Ejlli:2018hke}
\bibinfo{author}{\bibfnamefont{D.}~\bibnamefont{Ejlli}} \bibnamefont{and} \bibinfo{author}{\bibfnamefont{V.~R.} \bibnamefont{Thandlam}}, \bibinfo{journal}{Phys. Rev. D} \textbf{\bibinfo{volume}{99}}, \bibinfo{pages}{044022} (\bibinfo{year}{2019}), \eprint{1807.00171}.

\bibitem[{\citenamefont{{Astropy Collaboration} et~al.}(2022)\citenamefont{{Astropy Collaboration}, {Price-Whelan}, {Lim}, {Earl}, {Starkman}, {Bradley}, {Shupe}, {Patil}, {Corrales}, {Brasseur} et~al.}}]{astropy:2022}
\bibinfo{author}{\bibnamefont{{Astropy Collaboration}}}, \bibinfo{author}{\bibfnamefont{A.~M.} \bibnamefont{{Price-Whelan}}}, \bibinfo{author}{\bibfnamefont{P.~L.} \bibnamefont{{Lim}}}, \bibinfo{author}{\bibfnamefont{N.}~\bibnamefont{{Earl}}}, \bibinfo{author}{\bibfnamefont{N.}~\bibnamefont{{Starkman}}}, \bibinfo{author}{\bibfnamefont{L.}~\bibnamefont{{Bradley}}}, \bibinfo{author}{\bibfnamefont{D.~L.} \bibnamefont{{Shupe}}}, \bibinfo{author}{\bibfnamefont{A.~A.} \bibnamefont{{Patil}}}, \bibinfo{author}{\bibfnamefont{L.}~\bibnamefont{{Corrales}}}, \bibinfo{author}{\bibfnamefont{C.~E.} \bibnamefont{{Brasseur}}}, \bibnamefont{et~al.}, \bibinfo{journal}{\apj} \textbf{\bibinfo{volume}{935}}, \bibinfo{eid}{167} (\bibinfo{year}{2022}), \eprint{2206.14220}.

\bibitem[{\citenamefont{{Perley} and {Butler}}(2017)}]{2017ApJS..230....7P}
\bibinfo{author}{\bibfnamefont{R.~A.} \bibnamefont{{Perley}}} \bibnamefont{and} \bibinfo{author}{\bibfnamefont{B.~J.} \bibnamefont{{Butler}}}, \bibinfo{journal}{Astrophys. J. Suppl. Ser.} \textbf{\bibinfo{volume}{230}}, \bibinfo{eid}{7} (\bibinfo{year}{2017}), \eprint{1609.05940}.

\bibitem[{\citenamefont{Cowan et~al.}(2011)\citenamefont{Cowan, Cranmer, Gross, and Vitells}}]{Cowan:2010js}
\bibinfo{author}{\bibfnamefont{G.}~\bibnamefont{Cowan}}, \bibinfo{author}{\bibfnamefont{K.}~\bibnamefont{Cranmer}}, \bibinfo{author}{\bibfnamefont{E.}~\bibnamefont{Gross}}, \bibnamefont{and} \bibinfo{author}{\bibfnamefont{O.}~\bibnamefont{Vitells}}, \bibinfo{journal}{Eur. Phys. J. C} \textbf{\bibinfo{volume}{71}}, \bibinfo{pages}{1554} (\bibinfo{year}{2011}), \bibinfo{note}{[Erratum: Eur.Phys.J.C 73, 2501 (2013)]}, \eprint{1007.1727}.

\bibitem[{\citenamefont{Beggan}(2022)}]{BegganEvidencebased2022}
\bibinfo{author}{\bibfnamefont{C.~D.} \bibnamefont{Beggan}}, \bibinfo{journal}{Earth, Planets and Space} \textbf{\bibinfo{volume}{74}}, \bibinfo{pages}{17} (\bibinfo{year}{2022}), \urlprefix\url{https://doi.org/10.1186/s40623-022-01572-y}.

\bibitem[{\citenamefont{Tsyganenko and Andreeva}(2016)}]{https://doi.org/10.1002/2016JA023217}
\bibinfo{author}{\bibfnamefont{N.~A.} \bibnamefont{Tsyganenko}} \bibnamefont{and} \bibinfo{author}{\bibfnamefont{V.~A.} \bibnamefont{Andreeva}}, \bibinfo{journal}{Journal of Geophysical Research: Space Physics} \textbf{\bibinfo{volume}{121}}, \bibinfo{pages}{10,786} (\bibinfo{year}{2016}), \urlprefix\url{https://agupubs.onlinelibrary.wiley.com/doi/abs/10.1002/2016JA023217}.

\bibitem[{\citenamefont{Kelley}(2009)}]{kelley2009earth}
\bibinfo{author}{\bibfnamefont{M.~C.} \bibnamefont{Kelley}}, \emph{\bibinfo{title}{The Earth's ionosphere: Plasma physics and electrodynamics}}, vol.~\bibinfo{volume}{96} (\bibinfo{publisher}{Academic press}, \bibinfo{year}{2009}).

\end{thebibliography}




\onecolumngrid
\clearpage

\setcounter{page}{1}
\setcounter{equation}{0}
\setcounter{figure}{0}
\setcounter{table}{0}
\setcounter{section}{0}
\setcounter{subsection}{0}
\renewcommand{\theequation}{S.\arabic{equation}}
\renewcommand{\thefigure}{S\arabic{figure}}
\renewcommand{\thetable}{S\arabic{table}}
\renewcommand{\thesection}{\Roman{section}}
\renewcommand{\thesubsection}{\Alph{subsection}}
\newcommand{\ssection}[1]{
    \addtocounter{section}{1}
    \section{\thesection.~~~#1}
    \addtocounter{section}{-1}
    \refstepcounter{section}
}
\newcommand{\ssubsection}[1]{
    \addtocounter{subsection}{1}
    \subsection{\thesubsection.~~~#1}
    \addtocounter{subsection}{-1}
    \refstepcounter{subsection}
}
\newcommand{\fakeaffil}[2]{$^{#1}$\textit{#2}\\}

\setcounter{figure}{0}
\setcounter{table}{0}

\author{Hongliang Tian}
\email{hongliangtian@njnu.edu.cn}
\affiliation{Department of Physics and Institute of Theoretical Physics, Nanjing Normal University, Nanjing, 210023, China}

\author{Lei Wu}
\email{leiwu@njnu.edu.cn}
\affiliation{Department of Physics and Institute of Theoretical Physics, Nanjing Normal University, Nanjing, 210023, China}
\affiliation{Nanjing Key Laboratory of Particle Physics and Astrophysics, Nanjing, 210023, China}

\author{Xiaolong Yang}
\email{yangxl@shao.ac.cn}
\affiliation{Shanghai Astronomical Observatory, Chinese Academy of Sciences, Shanghai 200030, China}
\affiliation{Shanghai Key Laboratory of Space Navigation and Positioning Techniques, Shanghai 200030, China}

\author{Qiang Yuan}
\email{yuanq@pmo.ac.cn}
\affiliation{Key Laboratory of Dark Matter and Space Astronomy, Purple Mountain Observatory, Chinese Academy of Sciences, Nanjing 210023, China}
\affiliation{School of Astronomy and Space Science, University of Science and Technology of China, Hefei 230026, China}

\author{Bin Zhu}
\email{zhubin@mail.nankai.edu.cn}
\affiliation{School of Physics, Yantai University, Yantai 264005, China}

\thispagestyle{empty}
\begin{center}
    \begin{spacing}{1.2}
        \textbf{\large 
            \hypertarget{sm}{Supplemental Material:}
            \\Search for High-Frequency Gravitational Waves via Geomagnetic Conversion with Radio Telescopes
        }
    \end{spacing}
    \par\smallskip
    Hongliang Tian,$^{1}$
    Lei Wu,$^{1,2}$
    Xiaolong Yang$^{3,4}$
    Qiang Yuan$^{5,6}$
    and Bin Zhu,$^{7}$
    \par
    {\small
        \fakeaffil{1}{Department of Physics and Institute of Theoretical Physics, Nanjing Normal University, Nanjing, 210023, China}
        \fakeaffil{2}{Nanjing Key Laboratory of Particle Physics and Astrophysics, Nanjing, 210023, China} 
        \fakeaffil{3}{Shanghai Astronomical Observatory, Chinese Academy of Sciences, Shanghai 200030, China}
        \fakeaffil{4}{Shanghai Key Laboratory of Space Navigation and Positioning Techniques, Shanghai 200030, China}
        \fakeaffil{5}{Key Laboratory of Dark Matter and Space Astronomy, Purple Mountain Observatory, Chinese Academy of Sciences, Nanjing 210023, China}
        \fakeaffil{6}{School of Astronomy and Space Science, University of Science and Technology of China, Hefei 230026, China}
        \fakeaffil{7}{Department of Physics, Yantai University, Yantai 264005, China}
        
    }

\end{center}
\par\smallskip

This Supplemental Material details the computational framework and sensitivity projections for the search for high-frequency gravitational waves via geomagnetic conversion. It begins with the theoretical derivation of the Gertsenshtein effect in the Earth's magnetosphere, including the implementation of the IGRF-14 model and the coordinate transformation from the ECEF to the local ENU frame. We then describe the data analysis pipeline applied to archival VLA and ALMA observations, covering RFI mitigation and source subtraction. Subsequently, we present the sensitivity projection for the Square Kilometre Array. Finally, we discuss the uncertainty estimates associated with the geomagnetic field modeling and plasma effects.\\

{\bf Gertsenshtein Effect.}
The conversion between gravitational waves and electromagnetic waves originates from the coupling of metric perturbations and the electromagnetic tensor. The Lagrangian density is written as
\begin{equation}
\begin{aligned}
\mathcal{L}_{0}\supset \sqrt{-g}\frac{R}{\kappa^{2}}-\frac{1}{4}\sqrt{-g}g_{\mu\nu}g_{\rho\sigma}F^{\mu\rho}F^{\nu\sigma} \;,
\end{aligned}
\end{equation}
where $R$ is Ricci scalar, $g_{\mu\nu}$ is the metric tensor\footnote{In this article $\eta_{\mu\nu}={\rm diag}(-1,1,1,1)$, and $g = {\rm det}(g_{\mu\nu})$.} , and $F^{\mu\nu}$ is the electromagnetic tensor. After expanding the metric perturbation $g_{\mu\nu}=\eta_{\mu\nu}+\kappa h_{\mu\nu}$, the coupling term between gravitational waves and the electromagnetic field is obtained by $\mathcal{L}_{\text{c}}=\frac{\kappa}{2}h_{\mu\nu}T^{\mu\nu}$, where $T^{\mu\nu}$ represents the electromagnetic energy-momentum tensor. In addition, the Lagrangian contains the Euler–Heisenberg contribution $\mathcal{L}_{\text{E-H}}=\frac{\alpha^{2}}{90 m_{\mathrm{e}}^{4}}\sqrt{-g}\left[(F_{\mu\nu}F^{\mu\nu})^{2}+\frac{7}{4}(F_{\mu\nu}\tilde{F}^{\mu\nu})^{2}\right]$, which arises from the vacuum polarization. Given that the wavelength of photons generated from gravitational-wave conversion is much shorter than the characteristic spatial scale of the external magnetic field, the Wentzel-Kramers-Brillouin approximation can be employed to solve the dynamical mixing equations of motion. Moreover, because the two polarization modes of the gravitational wave decouple completely from their corresponding electromagnetic wave modes and can be treated independently, the leading-order equation takes the form:
\begin{align}
\label{eqn:dynamic}
\left[\omega+i\partial_r+
\begin{pmatrix}
  0&  \Delta_M\\
  \Delta_M&  \Delta_{\gamma,\times(+)}
\end{pmatrix}
\right]
\begin{pmatrix}
    h_{\times(+)}\\A_{\times(+)}
\end{pmatrix}=0\;.
\end{align}
The matter term $\Delta_M$ characterizes the coupling strength of gravitational wave and electromagnetic wave, $\Delta_{M}=\frac{1}{2}\kappa B_{t}$, where $B_t$ represents the tangential component of the magnetic field as a function of position. Meanwhile, the diagonal term $\Delta_{\gamma,\times(+)}=\Delta_{\text{vac},\times(+)}+\Delta_{\text{pla}}+\Delta_{\text{CM}}=\frac{7(4)\alpha\omega}{90\pi}\left(\frac{B_{t}}{B_{c}}\right)^{2}-\frac{2\pi\alpha n_{\text{e}}}{\omega m_{\text{e}}}+\Delta_{\text{CM}}$ includes the quantum electrodynamics effect due to vacuum polarization, the plasma effect due to refraction of the photon in the medium, and the birefringence effect like the Cotton-Mouton effect. The notation $\times(+)$ represents two different polarizations of gravitational waves and photons, and $B_{c}=m_{\text{e}}^2/e$ is the critical magnetic field. To solve Eq.~\eqref{eqn:dynamic}, we employ the Dyson series expansion in the interaction picture, whose first-order term yields the conversion amplitude\cite{Raffelt:1987im}
\begin{equation}
    P=\left|\int_{l_{0}}^{l_{1}}d l \,\Delta_{M}(l)\exp{\left[-i\int_{l_{0}}^l d l' \Delta_{\gamma}(l')\right]}\right|^{2} \;.
    \label{eq:probability}
\end{equation}
As noted in the literature~\cite{Liu:2023mll,Ejlli:2018hke}, the plasma effect, the Cotton-Mouton term, and the vacuum polarization are all negligible for gravitational wave--electromagnetic wave conversion in the geomagnetic field over the frequency range considered here. Consequently, We have $\Delta_{\gamma} = 0$ for both modes. Applying these conditions to Eq.~\eqref{eq:probability} for a specific direction defined by the zenith $\theta$ and the azimuth $\phi$, we obtain Eq.~\eqref{eqn:probability}. 

{\bf IGRF Model.}
The International Geomagnetic Reference Field model (IGRF-14 is adopted in this study) provides a time-dependent representation of the geomagnetic field via spherical harmonic coefficients spanning from 1900 to 2025, along with a predictive component for subsequent years. Widely utilized in geophysics, space science, and planetary studies, the model offers a standardized reference for describing the global geomagnetic structure and its secular variation. The specific description of the geomagnetic field is the scalar potential function $V_{0}$ given by
\begin{equation}
    V_{0}(r,\lambda,\psi)=\sum_{l=1}^{\infty}\sum_{m=0}^{l}\frac{r_{0}^{l+2}}{r^{l+1}}(g_{l m}\cos{m\psi}+h_{l m}\sin{m\psi})P_{l}^{m}(\sin{\lambda}),
\end{equation}
with the reference Earth radius $r_{\oplus} = 6371.2$ km. We denote the geodetic latitude and longitude of the observation station as $\lambda$ and $\psi$, respectively. In the above formula, $P_{l}^{m}$ is the Schmidt-normalized associated Legendre polynomial defined as
\begin{equation}
    P_{l}^{m}(x)=\sqrt{(2-\delta_{0}^{m})\frac{(l-m)!}{(l+m)!}}(1-x^{2})^{m/2}\frac{d^{m}}{d\,x^{m}}P_{l}(x),
\end{equation}
where $P_{l}(x)$ are the Legendre polynomial. We connect the scalar potential with the magnetic induction by $B=-\nabla\,V$. The IGRF model provides the Gauss coefficients $g_{l m}$ and $h_{l m}$ for $l\leq 13$ at five-year intervals and their values at intermediate time can be calculated by linear interpolation. Leveraging the comprehensive IGRF model, we can map the directional conversion probability of gravitational waves into electromagnetic waves across space and time. 

It is particularly noted that the IGRF model is referenced to the global ECEF coordinate system. A coordinate transformation from the ECEF system to the ENU system is therefore required for the computation. We proceed to establish the general transformation from the local tangent plane, defined by the ENU system with origin $O_{2}$, to the ECEF system with origin $O_{1}$. The point $A$ in the ENU system is described by the radial distance $r$, zenith angle $\theta$, and azimuth angle $\phi$. The zero point of the azimuth is defined as the north direction, with $90^{\circ}$ corresponding to the east. We first convert the spherical coordinates in the ENU system to their Cartesian components along each axis, that is $\overrightarrow{O_{2}A}=r\sin{\theta}\cos{\phi}\,\widehat{e_{\text{N}}}+r\sin{\theta}\sin{\phi}\,\widehat{e_{\text{E}}}+r\cos{\theta}\,\widehat{e_{z'}}$
where $\widehat{e_{\text{N}}}, \widehat{e_{\text{E}}}$ and $\widehat{e_{\text{z}'}}$ are the unit vectors toward north, east and zenith. The unit vectors of spherical coordinate system $O_{2}$ resolved in terms of the Cartesian coordinate system $O_{1}$ is
\begin{equation}
\begin{pmatrix}
 \widehat{e_{x}}\\
 \widehat{e_{y}}\\
 \widehat{e_{z}}
\end{pmatrix}
=
\begin{pmatrix}
  \cos{\lambda}\cos{\psi}& \sin{\lambda}\cos{\psi} & -\sin{\psi}\\
  \cos{\lambda}\sin{\psi}& \sin{\lambda}\sin{\psi} & \cos{\psi}\\
  \sin{\lambda}& -\cos{\lambda} & 0
\end{pmatrix}
\begin{pmatrix}
 \widehat{e_{r}} \\
 \widehat{e_{\theta}}\\
 \widehat{e_{\phi}}
\end{pmatrix},
\end{equation}
and for unit vectors $\widehat{e_{\text{N}}}=-\widehat{e_{\theta}}, \widehat{e_{\text{E}}}=\widehat{e_{\phi}}$ and $\widehat{e_{\text{z}'}}=\widehat{e_{r}}$. Thus we express $\overrightarrow{O_{2}A}$
as
\begin{equation}
\begin{aligned}
\overrightarrow{O_{2}A}=r\,&(\cos{\theta}\cos{\lambda}\cos{\psi}-\sin{\theta}\sin{\phi}\sin{\psi}-\sin{\theta}\cos{\phi}\sin{\lambda}\cos{\psi},\\
&\sin{\theta}\sin{\phi}\cos{\psi}+\cos{\theta}\cos{\lambda}\sin{\psi}-\sin{\theta}\cos{\phi}\sin{\lambda}\sin{\psi},\\
&\sin{\theta}\cos{\phi}\cos{\lambda}+\cos{\theta}\sin{\lambda}).
\end{aligned}
\end{equation}
$\overrightarrow{O_{1}O_{2}}$ is straightforward and can be obtained directly $\overrightarrow{O_{1}O_{2}}=r_{0}\,(\cos{\lambda}\cos{\psi},\cos{\lambda}\sin{\psi},\sin{\lambda})$. 
Hence, the coordinates of A can be written as $\overrightarrow{O_{1}A}=\overrightarrow{O_{1}O_{2}}+\overrightarrow{O_{2}A}$.

Indeed, we aim to calculate the magnitude of the transverse magnetic field component perpendicular to the propagation direction of gravitational waves. To achieve this, we employ the Schmidt orthogonalization method. The magnitude of the transverse magnetic field can be expressed as
\begin{equation}
    B_{t}(A)=\left|\overrightarrow{B}(A)-\frac{\overrightarrow{B}(A) \cdot \overrightarrow{O_{2}A}}{\left|\overrightarrow{O_{2}A}\right|^{2}}\,\overrightarrow{O_{2}A}\right|,
\end{equation}
which relates to Eq.~\eqref{eqn:probability}.

\begin{figure}[ht!]
    \centering
    \includegraphics[width=0.6\linewidth]{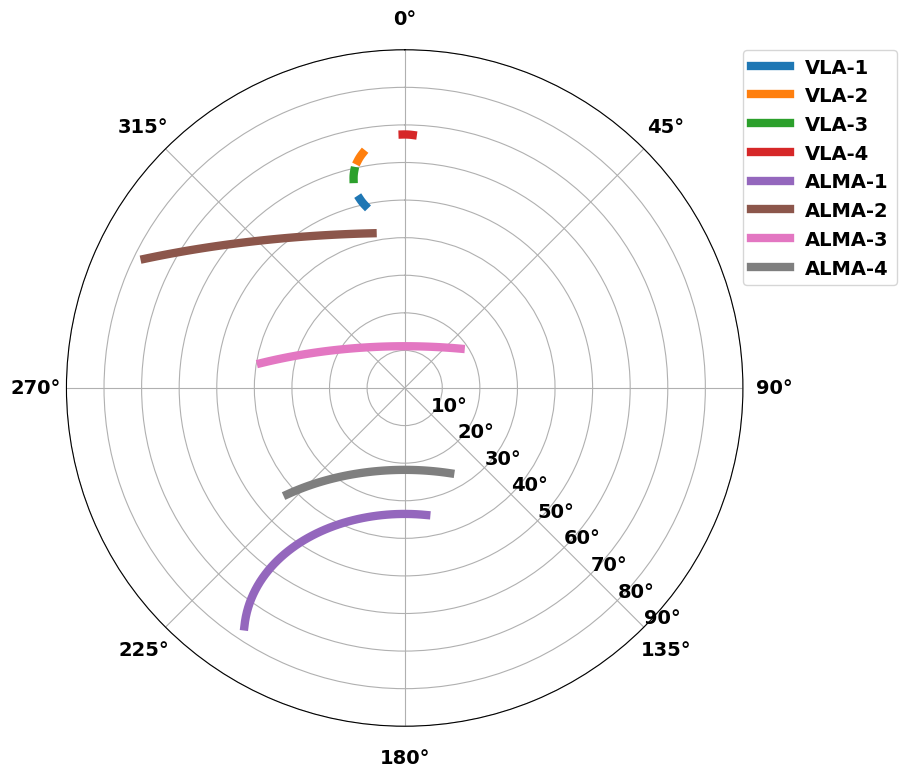}
    \caption{The ENU-frame trajectories of target sources observed by the VLA and the ALMA. The trajectories are computed with the {\tt Astropy}\cite{astropy:2022} tool, for the geographic coordinates of the ALMA ($-23.02^\circ,~-67.75^\circ,~5000~\mathrm{m}$) and the VLA ($34.08^\circ,~-107.62^\circ,~2124~\mathrm{m}$).}
    \label{fig:trajectory}
\end{figure}

The electromagnetic wave signal considered in this work depends on the geometry of the geomagnetic field, and hence the conversion probability shown in Figure \ref{fig:all_sky_map} is presented in the ENU coordinate. However, when the telescope tracks a celestial source with fixed coordinate (Right Ascension and Declination), the projection of that position on the celestial sphere shifts over time due to the Earth’s rotation, giving a curved trajectory in the ENU coordinate. Here we show in  Figure \ref{fig:trajectory} with the trajectories of 8 selected observations, 4 for the VLA and 4 for the ALMA, in the ENU coordinate. The basic information of the 8 observations is given in Table \ref{tab:trajectories_data}. 

\begin{table*}[ht!]
 \centering
 \begin{tabular}{c|cc|c|c}
  \hline\hline
  Station & \multicolumn{2}{c|}{Coordinate of FoV center} & Start epoch & Exposure \\  
  number &  Right Ascension (RA) & ~~Declination (DEC) & UT & (hour)\\
  \hline
  VLA-1 & $18^h00^m45^s.684$ & ~~$+78^{\circ}28'04''.018$  & ~~2025-04-28T14:18:15.000~~ & 0.70\\
  VLA-2 & $18^h00^m45^s.684$ & ~~$+78^{\circ}28'04''.018$  & ~~2025-04-27T18:34:18.000~~ & 0.70\\
  VLA-3 & $18^h00^m45^s.684$ & ~~$+78^{\circ}28'04''.018$  & ~~2025-04-21T17:20:45.000~~ & 0.70\\
  VLA-4 & $18^h00^m45^s.684$ & ~~$+78^{\circ}28'04''.018$  & ~~2025-04-14T23:27:44.000~~ & 0.70\\
  ALMA-1 & $23^h49^m58^s.012$ & ~~$-56^{\circ}40'17''.779$  & ~~2024-06-24T09:30:47.232~~ & 7.33\\ 
  ALMA-2 & $04^h31^m47^s.983$ & ~~$+18^{\circ}11'42''.601$  & ~~2023-10-14T08:06:42.384~~ & 3.90\\ 
  ALMA-3 & $13^h47^m38^s.556$ & ~~$-11^{\circ}47'17''.093$  & ~~2024-03-15T05:46:06.912~~ & 3.61\\ 
  ALMA-4 & $20^h41^m10^s.156$ & ~~$-44^{\circ}53'31''.667$  & ~~2024-05-02T09:23:43.440~~ & 4.07\\ 
  \hline
 \end{tabular}
 \caption{Information of the selected 8 observations of the VLA and the ALMA for illustrating trajectories in Figure~\ref{fig:trajectory}.}
 \label{tab:trajectories_data}
\end{table*}

{\bf Data Analysis.}
The archival data from the VLA and the ALMA facilities are used in this study. All datasets are retrieved from the NRAO Data Archive\footnote{\url{https://data.nrao.edu/}} and the ALMA Science Archive\footnote{\url{https://almascience.nrao.edu/}} for manual calibration and processing. Standard flagging procedures were applied to remove the RFI. 
Note that there are some temporal variabilities of the RFI, due to e.g. the variation of the interfering source, the varying antenna susceptibility, the target source location, and the satellite transmission.
The calibration is done following the procedures outlined in the {\tt CASA} cookbook, including ionospheric corrections with data from the CDDIS archive\footnote{\url{https://cddis.nasa.gov/}}. The flux density scales for primary calibrators are set using the Perley-Butler 2017 standard\cite{2017ApJS..230....7P}. Complex gain solutions derived from secondary calibrators are then transferred to the target fields. 

We perform imaging using the {\tt tclean} task in {\tt CASA} in frequency mode to generate three-dimensional spectral image cubes, wherein the deconvolution is executed independently for each spectral channel. We conduct the full-field imaging to map all emission within the primary beam, which ensures the identification of bright, compact in-field sources used to construct models for self-calibration. Then we apply several iterations of the phase-only self-calibration, followed by the amplitude-and-phase calibration, to mitigate atmospheric phase errors and maximize the dynamic range of the final images. To recover emission across a broad range of spatial scales, we employ a multi-scale deconvolver algorithm. The scales are strategically selected to correspond to the synthesized beam size and several larger multiples thereof, ensuring the recovery of both compact and extended structures. 

The resulting data cubes, comprising two spatial dimensions (Right Ascension and Declination) and one spectral dimension (frequency), provide a comprehensive mapping of the radio emission. 
For each spectral channel in the image cube, we generate a source map by employing the \textsc{PyBDSF} (Python Blob Detection and Source Finder)\footnote{\url{https://pybdsf.readthedocs.io}}, which is developed for radio source detection and extraction in the image domain. We calculate the local root-mean-square (RMS) noise using an adaptive sliding box to account for non-uniform noise distributions across the primary beam. Source detection is performed where significant source components are identified at a threshold of ($5\sigma$) relative to the local RMS noise. The full extent of each source is then defined as the contiguous emission region enclosed by the ($3\sigma$) contour, which is adopted as the island boundary.
We employ the $\text{À trous}$ wavelet decomposition module within the \textsc{PyBDSF} (specifically setting $atrous\_jmax=15$) to effectively model and subtract discrete astrophysical foregrounds. This multi-order approach is designed to identify and deconvolve compact points and moderately extended radio sources characterized by high-frequency spatial gradients. While the wavelet decomposition handles structures up to the specified $jmax$ scale, the target HFGWs signal is characterized by a significantly smoother, large-scale morphology that lacks sharp edges or high-surface-brightness peaks associated with modeled source components. The difference between the original data and the modeled astrophysical sources is taken as a residual map for the HFGW-induced signal search. This procedure ensures that the diffuse signal of interest is preserved in the residual while contamination from point-like and multi-scale galactic/extragalactic emission is suppressed.

\begin{figure}[htp!]
    \centering
    \includegraphics[width=1.0\linewidth]{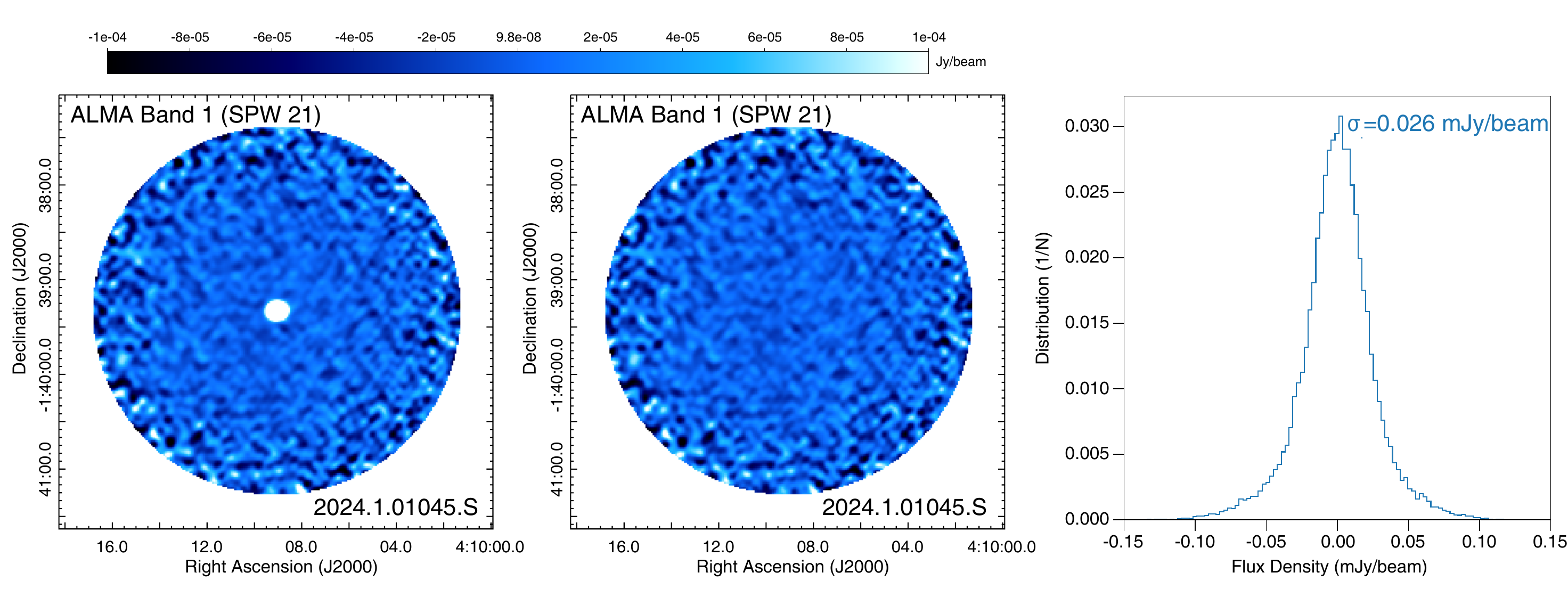}
    \caption{Left: calibrated image of one observation with the ALMA (central frequency of 37.40\,GHz and bandwidth of 1.47\,GHz). Middle: residual image after subtracting point-like and extended sources. Right: one-dimensional histogram distribution of the flux density for each beam.}
    \label{fig:sigma}
\end{figure}

\begin{figure}[htp!]
    \centering
    \includegraphics[width=0.8\linewidth]{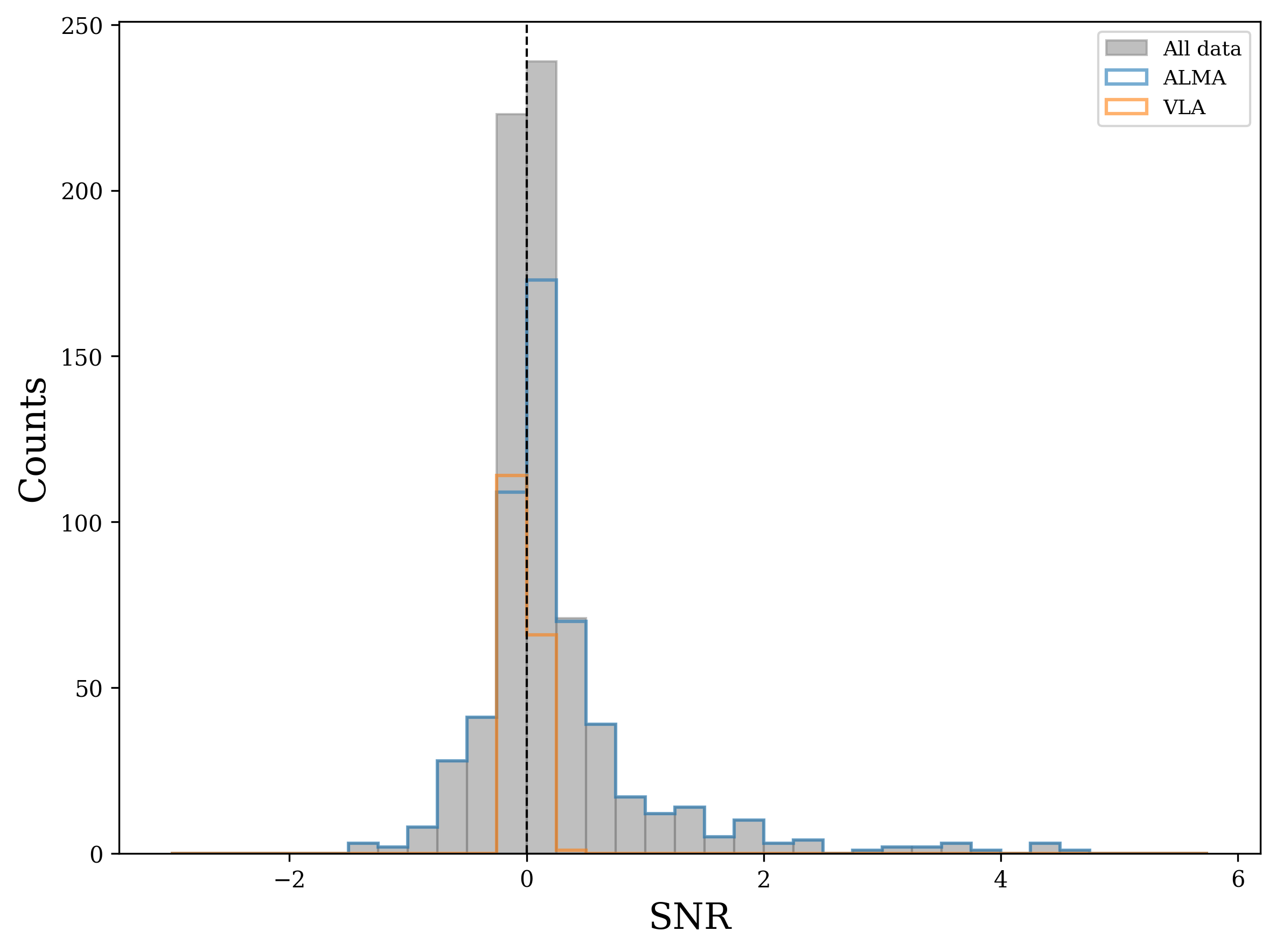}
    \caption{Histogram of the signal-to-noise ratio (SNR) for all data used in our analysis. SNR is defined as the ratio of the mean noise to its standard deviation. The grey filled histogram shows the combined distribution for all stations, while the blue and orange step histograms correspond to ALMA and VLA, respectively. The bin width is 0.25, and the vertical dashed line marks SNR = 0.}
    \label{fig:snr}
\end{figure}

Since the bandwidth of each spectral channel is very narrow, we rebin the spectral channels properly. The residual map is then integrated to give the total flux density in the FoV. To obtain the RMS of the total emission in the FoV, we employ an error propagation of the RMS per beam, $\sigma_{\rm FoV}=\sqrt{N_{\rm b}}\sigma_{\rm b}$, where $\sigma_{\rm b}$ is the RMS per beam (width of the histogram shown in Figure \ref{fig:sigma}) and $N_{\rm b}$ is the number of beam in the FoV. We find that the absolute values of the total flux densities are typically smaller than the RMS values, suggesting that no significant residual signal exists in the data. The signal-to-noise ratio (SNR) distribution of the data is shown in Figure~\ref{fig:snr}. One can see that no observation has SNR bigger than $5$. The distribution of the SNR is slightly asymmetric, with a tail towards the positive end. This might be due to  faint residual emission in the FoV not fully removed by the processing procedure. 
Several rows of the complete data are shown in Table \ref{tab:data}, and the complete data file is publicly available~\href{https://github.com/Hong-liangTian/Supplementary-Data-File}{\faGithub}.

\afterpage{
\begin{table}[htp]
\centering
\caption{Summary of the results of the data used in this work.}
\label{tab:data}
\footnotesize
\setlength{\tabcolsep}{3pt}
\begin{tabular}{ccccccccccc}
\toprule
\hline
Station & RA & DEC & Start epoch & Exposure & $\theta_{\text{FoV}}$ & $\nu_c$ & $\Delta\nu$ & Flux & RMS & UL$_{95\%}$\\
\hline
        &  &  &  & (hour) & (degree) & (GHz) & (GHz) & (mJy) & (mJy) & (Jy/sr) \\
\midrule
\hline
VLA & $18^h00^m45^s.6839$ & $+78^{\circ}28'04''.018$ &2025-04-28T14:18:15.000  &0.702  &  &  &  &  &  &  \\
VLA & $18^h00^m45^s.6839$ & $+78^{\circ}28'04''.018$ &2025-04-27T18:34:18.000  &0.702  &  &  &  &  &  &  \\
VLA & $18^h00^m45^s.6839$ & $+78^{\circ}28'04''.018$ &2025-04-21T17:20:45.000  &0.702  &  &  &  &  &  &  \\
VLA & $18^h00^m45^s.6839$ & $+78^{\circ}28'04''.018$ &2025-04-14T23:27:44.000  &0.702  &  &  &  &  &  &  \\
VLA & $18^h00^m45^s.6839$ & $+78^{\circ}28'04''.018$ &2025-03-17T21:56:45.000  &0.702  &  &  &  &  &  &  \\
VLA & $18^h00^m45^s.6839$ & $+78^{\circ}28'04''.018$ &2025-03-10T22:13:43.000  &0.702  &  &  &  &  &  &  \\
VLA & $18^h00^m45^s.6839$ & $+78^{\circ}28'04''.018$ &2024-01-12T02:38:33.000 &0.702 & 0.670 & 1.026 & 0.064 & 0.00 & 0.84 & $1.53\times10^{1}$ \\
VLA & $18^h00^m45^s.6839$ & $+78^{\circ}28'04''.018$ &2024-01-02T20:13:11.000  &0.702  &  &  &  &  &  &  \\
VLA & $18^h00^m45^s.6839$ & $+78^{\circ}28'04''.018$ &2023-12-11T18:04:07.000  &0.702  &  &  &  &  &  &  \\
VLA & $18^h00^m45^s.6839$ & $+78^{\circ}28'04''.018$ &2023-12-04T23:12:37.000  &0.702  &  &  &  &  &  &  \\
VLA & $18^h00^m45^s.6839$ & $+78^{\circ}28'04''.018$ &2023-11-21T22:18:22.000  &0.702  &  &  &  &  &  &  \\
VLA & $18^h00^m45^s.6839$ & $+78^{\circ}28'04''.018$ &2023-10-26T20:56:38.000  &0.702  &  &  &  &  &  &  \\
VLA & $18^h00^m45^s.6839$ & $+78^{\circ}28'04''.018$ &2023-10-25T18:19:24.000  &0.702  &  &  &  &  &  &  \\
\hline
VLA & $18^h00^m45^s.6839$ & $+78^{\circ}28'04''.018$ &2025-04-28T14:18:15.000  &0.702  &  &  &  &  &  &  \\
VLA & $18^h00^m45^s.6839$ & $+78^{\circ}28'04''.018$ &2025-04-27T18:34:18.000  &0.702  &  &  &  &  &  &  \\
VLA & $18^h00^m45^s.6839$ & $+78^{\circ}28'04''.018$ &2025-04-21T17:20:45.000  &0.702  &  &  &  &  &  &  \\
VLA & $18^h00^m45^s.6839$ & $+78^{\circ}28'04''.018$ &2025-04-14T23:27:44.000  &0.702  &  &  &  &  &  &  \\
VLA & $18^h00^m45^s.6839$ & $+78^{\circ}28'04''.018$ &2025-03-17T21:56:45.000  &0.702  &  &  &  &  &  &  \\
VLA & $18^h00^m45^s.6839$ & $+78^{\circ}28'04''.018$ &2025-03-10T22:13:43.000  &0.702  &  &  &  &  &  &  \\
VLA & $18^h00^m45^s.6839$ & $+78^{\circ}28'04''.018$ &2024-01-12T02:38:33.000 &0.702 & 0.621 & 1.106 & 0.096 & 0.00 & 2.50 & $5.31\times10^{1}$ \\
VLA & $18^h00^m45^s.6839$ & $+78^{\circ}28'04''.018$ &2024-01-02T20:13:11.000  &0.702  &  &  &  &  &  &  \\
VLA & $18^h00^m45^s.6839$ & $+78^{\circ}28'04''.018$ &2023-12-11T18:04:07.000  &0.702  &  &  &  &  &  &  \\
VLA & $18^h00^m45^s.6839$ & $+78^{\circ}28'04''.018$ &2023-12-04T23:12:37.000  &0.702  &  &  &  &  &  &  \\
VLA & $18^h00^m45^s.6839$ & $+78^{\circ}28'04''.018$ &2023-11-21T22:18:22.000  &0.702  &  &  &  &  &  &  \\
VLA & $18^h00^m45^s.6839$ & $+78^{\circ}28'04''.018$ &2023-10-26T20:56:38.000  &0.702  &  &  &  &  &  &  \\
VLA & $18^h00^m45^s.6839$ & $+78^{\circ}28'04''.018$ &2023-10-25T18:19:24.000  &0.702  &  &  &  &  &  &  \\
\multicolumn{11}{c}{$\cdots$} \\
\hline
ALMA & $04^h10^m18^s.1473$ & $-01^{\circ}41'36''.278$ & 2024-11-02T05:24:28.128 & 1.46 & 0.038 & 39.23 & 1.68 & 3.56 & 0.75 & $1.46\times10^{4}$ \\
ALMA & $21^h01^m30^s.9783$ & $-60^{\circ}51'01''.861$ & 2024-03-11T15:41:23.664 & 1.46 & 0.035 & 39.31 & 1.77 & -0.02 & 1.22 & $8.14\times10^{3}$ \\
ALMA & $23^h49^m58^s.0116$ & $-56^{\circ}40'17''.779$ & 2024-06-24T09:30:47.232 & 7.33 & 0.071 & 39.99 & 1.81 & 0.18 & 4.28 & $7.11\times10^{3}$ \\
ALMA & $21^h01^m30^s.9783$ & $-60^{\circ}51'01''.861$ & 2024-03-11T15:41:23.664 & 1.46 & 0.035 & 41.29 & 1.70 & -0.34 & 1.10 & $6.63\times10^{3}$ \\
ALMA & $20^h41^m10^s.1555$ & $-44^{\circ}53'31''.667$ & 2024-05-02T09:23:43.440 & 4.07 & 0.065 & 42.52 & 1.78 & -0.05 & 1.65 & $3.16\times10^{3}$ \\
\multicolumn{11}{c}{$\cdots$} \\
\bottomrule
\hline
\end{tabular}
\end{table}
}

{\bf Statistical Method.}

Assuming that the measurement follows a Gaussian distribution, we adopt a likelihood-based test statistic~\cite{Cowan:2010js} for a single frequency bin as
\begin{equation}
q(I)=\left\{
\begin{matrix}
 \frac{(I - \hat{I})^2}{\sigma^2},\qquad I \geq \hat{I}\\
 0,\qquad \qquad ~~I<\hat{I}
\end{matrix}\right.,
\end{equation}
where $I$ is the hypothetical signal flux density, $\hat{I}$ is the best-fit flux density which is identical to the measured value of $I_0$ in our case. The test statistic $q$ follows a half-$\chi^2$ distribution with one degree of freedom
\begin{equation}
    f(q)=\frac{1}{2}\delta(q)+\frac{1}{2}\frac{1}{\sqrt{2\pi}}\frac{1}{\sqrt{q}}e^{-q/2},
\end{equation}
where $\delta(q)$ is the Dirac delta function. To obtain the one-sided upper limit, we employ a $p$-value defined as~\cite{Cowan:2010js}
\begin{equation}
p(I) = \frac{1 - \Phi\bigl(\sqrt{q(I)}\bigr)}{1 - \Phi\bigl(\sqrt{q(0)}\bigr)},
\label{eqn:p_value}
\end{equation}
where $\Phi$ is the cumulative distribution of the standard normal distribution. 

For a general observed flux $I_0$, the 95\% confidence level upper limit is obtained by solving $p(I)=0.05$, which yields
\begin{equation}
I^{95\%} = \hat{I} + z\sigma,
\qquad
z = \Phi^{-1}\Bigl[\,1 - 0.05\bigl(1-\Phi(\sqrt{q(0)})\bigr)\Bigr].
\end{equation}
When $I_0 \ge 0$, $q(0)=0$ and the denominator of Eq.~\ref{eqn:p_value} equals to $0.5$, leading to $z = 1.96$. For negative $I_0$, $z$ is larger than $1.96$. The upper limit on the characteristic strain is then
\begin{equation}
h_c^{95\%} = \sqrt{ \frac{\kappa^2 I^{95\%}}{\pi f \overline{P}}}.
\end{equation}
The resulting limits, computed separately for each frequency bin, are shown in Fig.~\ref{fig:hc_sensitivity}.


{\bf Projection of SKA.}
The expected signal power received by the telescope is $W=F\cdot A_{\text{eff}}\cdot \Delta f$ where $F$ is the radio flux density of signal, $A_{\text{eff}}$ is the effective area of radio telescope, and $\Delta f$ is the frequency bandwidth. The system noise power is $W_{n}=2kT_{\text{sys}}\cdot \Delta f$ where $k$ is Boltzmann constant and $T_{\text{sys}}$ is the equivalent noise temperature. 
The standard deviation of noise power per beam is $\sigma_W=\frac{W_{n}}{\sqrt{\Delta f\cdot t}}=2k T_{\text{sys}} \cdot \sqrt{\frac{\Delta f}{t}}$, with $t$ being the integration time of the observation. The estimated noise of flux density per beam is
\begin{equation}
    \sigma_{\text{beam}}=\frac{2k}{A_{\text{eff}}/T_{\text{sys}}}\sqrt{\frac{1}{\Delta f \cdot t}}.
\end{equation}
For the whole observational FoV, one has 
\begin{equation}
    \sigma_{\text{FoV}}=\sigma_{\text{beam}}\sqrt{\frac{\Omega_{\text{FoV}}}{\Omega_{\text{beam}}}},
\end{equation}
where $\Omega_{\rm FoV}/\Omega_{\rm beam}$ is the ratio of the solid angles of the FoV to the beam. 

We consider the following benchmark parameters of the SKA: the sensitivity $A_{\text{eff}}/T_{\text{sys}}$ from Ref.~\cite{Braun:2019gdo}, the bandwidth $\Delta f=0.1f$, the beam size $\theta_{\text{beam}}=(0.2~\mathrm{GHz}/f)~\mathrm{deg}$, the FoV size $\theta_{\text{FoV}}=36~\theta_{\text{beam}}$ which meets the parameter setting with VLA, and the integration time $t = 1\,\mathrm{h}$. The 95\% confidence level sensitivity on the characteristic strain $h_c$ derived from the estimated noise level of the SKA is shown by dashed line in Fig.~\ref{fig:hc_sensitivity}. Increasing the observational time will further improve the sensitivity on $h_c$ with the scaling form of $t^{-1/4}$. Note that our estimate is based on the incoherent summing mode of the array. A coherent approach would likely yield better sensitivity.

{\bf Systematic uncertainties}
The systematic uncertainties of our final constraints on the characteristic strain $h_c$ may originate from uncertainties in the geomagnetic field model, the decoherence effect from the ionosphere, and the systematic uncertainties in the radio flux density measurements.

\begin{figure}[htbp!]
    \centering
    \includegraphics[width=0.9\linewidth]{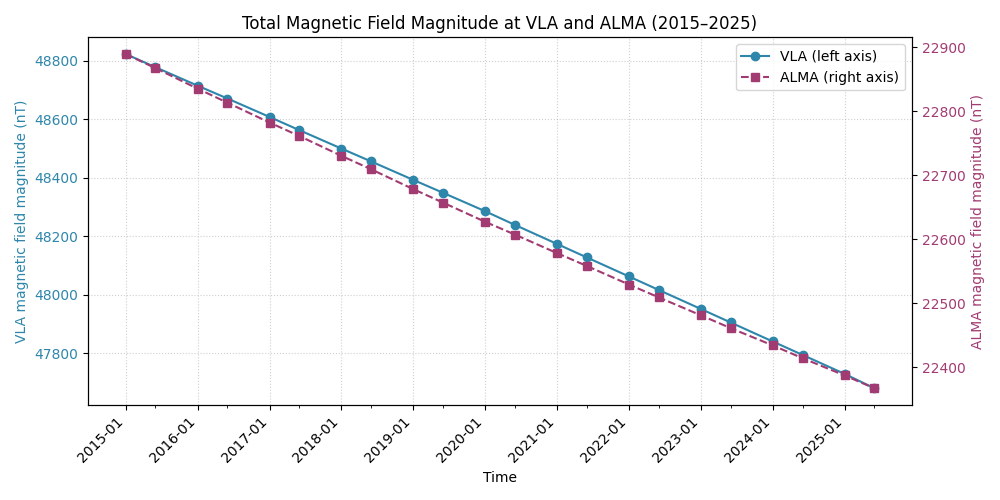}
    \caption{Temporal variations of the total magnetic field strength at the VLA (left axis) and the ALMA (right axis) sites from 2015 to 2025. Both results exhibit a gradual secular decline over the decade‑long time period.}
    \label{fig:geo}
\end{figure}

There are temporal variations of the geomagnetic field during the observations, as shown in Figure \ref{fig:geo}. This leads to about $2.3\%$ change of the upper limits of $h_{c}$. The discrepancy between the IGRF model predictions and the local magnetic field measurements is on average $\sim 10^3$ nT \cite{BegganEvidencebased2022}, corresponding to a relative uncertainty of $0.46\%$ in the characteristic strain $h_{c}$. Furthermore, the effect of solar winds on the magnetosphere at several Earth radii gives a variation of the magnetic field strength up to $40~\mathrm{nT}$ during the quiet phase and up to $300~\mathrm{nT}$ during strong activities \cite{https://doi.org/10.1002/2016JA023217}. This introduces an uncertainty of $\lesssim2\%$ in the characteristic strain $h_{c}$.
    
The electrons in the ionosphere would disrupt the coherence of the conversion between gravitational and electromagnetic waves~\cite{kelley2009earth}, depending on the electron density distribution and the gravitational wave frequency. Figure \ref{fig:survival} shows the normalized conversion probability as a function of frequency during the day time and the night time, respectively. It shows that due to the disruption of the coherence, the conversion probability swings between 1 GHz to 10 GHz, with two dips. Above 30 GHz, the probability keeps to be higher than 90\% since the plasma effect becomes small. Even for the low frequency region ($<30$ GHz), the maximum change of the conversion probability due to decoherence is smaller than a factor of $100$, which leads to weakening of the constraint on $h_{c}$ by less than a factor of 10.

As for the radio observations, a systematic uncertainty of 5\% to 10\% is inherently present for the VLA \footnote{\url{https://science.nrao.edu/facilities/vla/docs/manuals/oss2023B}} and the ALMA \footnote{{\url{https://almascience.nrao.edu/documents-and-tools/latest/alma-technical-handbook}}}, primarily arising from the absolute flux calibration. This translates to a relative uncertainty of 2.5\% to 5\% in the derived characteristic strain $h_c$. 



\begin{figure}[htbp!]
    \centering
    \includegraphics[width=0.7\linewidth]{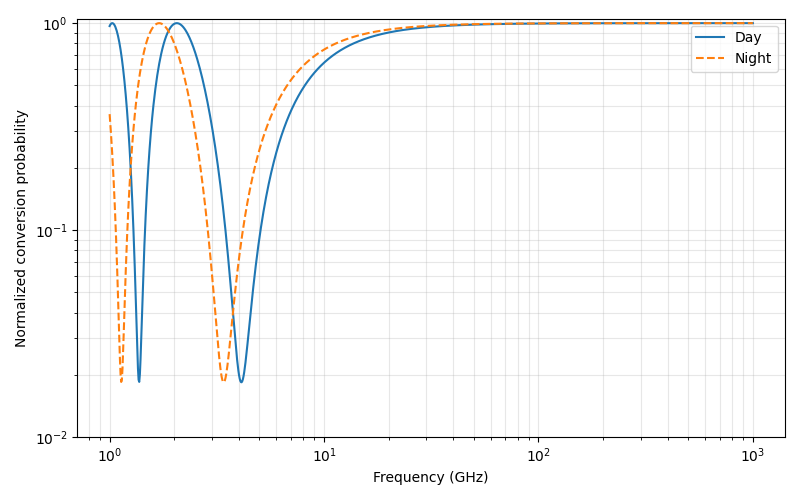}
    \caption{Normalized conversion probability as a function of frequency, for the day time (blue solid line) and night time (orange dashed line)  ionospheric electron density profiles~\cite{kelley2009earth}. Between 1 GHz and 10 GHz, the probability oscillates to produce two dips due to the decoherence. These minima occur where the phase mismatch from the plasma term largely cancels the conversion amplitude. Above 30 GHz, the plasma effect becomes very small and the coherence can be well satisfied.}
    \label{fig:survival}
\end{figure}

\begin{description}

\item[Data Availability] 
The archival VLA and ALMA data used in this study are publicly available from the NRAO Data Archive (\url{https://data.nrao.edu/}) and the ALMA Science Archive (\url{https://almascience.nrao.edu/}), respectively. Other derived data, supporting the findings of this study, are publicly available at \url{https://github.com/Hong-liangTian/Supplementary-Data-File}.

\item[Code Availability] 
Calculations regarding the generation of all‑sky conversion probability maps at specific terrestrial locations and the average conversion probability along specific trajectories, as described in Methods, are performed using Python. Annotated code is available from the corresponding author upon request.

\end{description}



\end{document}